\documentclass[english,aps,twocolumn,prd,superscriptaddress,showpacs,preprintnumbers,amsmath,amssymb]{revtex4-1}
\usepackage{amsmath,amssymb}
\usepackage{graphicx}
\usepackage{rotating}
\usepackage{dsfont}
\usepackage{color}
\usepackage[dvips]{epsfig}     
\usepackage{multirow}
\definecolor{refcol}{RGB}{0,100,170}
\usepackage{microtype}
\usepackage[colorlinks,linkcolor=refcol,citecolor=refcol,urlcolor=refcol]{hyperref}
\usepackage{breakurl}

\usepackage[utf8]{inputenc}

\graphicspath{{./figures/}}

\def\di{\displaystyle}

\def\bg{\begin{eqnarray}\begin{array}{rcl}\displaystyle}
\def\eg{\end{array} &\di    &\di   \end{eqnarray}}
\def\bm#1{\begin{eqnarray}\begin{array}{#1}\di}
\def\bmo#1{\begin{eqnarray*}\begin{array}{#1}\di}
\def\bml#1#2{\begin{eqnarray}\begin{array}{#1}\label{#2}\di}
\def\bgo{\begin{eqnarray*}\begin{array}{rcl}\displaystyle}
\def\ego{\end{array} &\di    &\di \nonumber  \end{eqnarray*}}

\def\btensor#1#2{\renew\left#1\begin{array}{#2}\di}
\def\brtensor#1#2#3{\ren#3\left#1\begin{array}{#2}}
\def\botensor#1#2{\renew\left#1\begin{array}{#2}}
\def\etensor#1{\end{array}\right#1}

\def\eq#1{(\ref{#1})}
\def\Eq#1{Eq.~(\ref{#1})}
\def\Fig#1{Fig.~\ref{#1}}

\def\det{{\rm det}}

\def\s0#1#2{\mbox{\small{$ \frac{#1}{#2} $}}}
\def\0#1#2{\frac{#1}{#2}}

\def\eq#1{(\ref{#1})}

\def\Eq#1{Eq.~(\ref{#1})}
\def\Fig#1{Fig.~\ref{#1}}
\def\Sec#1{Sec.~\ref{#1}}
\def\App#1{App.~\ref{#1}}
\def\Tab#1{Tab.~\ref{#1}}

\def\ren#1{\renewcommand{\arraystretch}{#1}}

\def\renew{\renewcommand{\arraystretch}{1}}

\definecolor{blue}{rgb}{0,0,1}

\definecolor{green}{rgb}{0,1,0}

\definecolor{red}{rgb}{1,0,0}

\definecolor{bjcol}{rgb}{1,.44,0.13}

\definecolor{fabcol}{rgb}{0.56,0.00,1.00}

\newcommand{\Tr}{\mathrm{Tr}}

\newcommand{\tr}{\mathrm{tr}}

\newcommand{\be}{\begin{eqnarray}}
\newcommand{\ee}{\end{eqnarray}}

\usepackage{babel}
\makeatother

\begin{document}

\title{Strangeness Neutrality and QCD Thermodynamics}

\author{Wei-jie Fu}
\affiliation{School of Physics , Dalian University of Technology, Dalian, 116024, P.R. China}

\author{Jan M. Pawlowski}
\affiliation{Institut
 f\"{u}r Theoretische Physik, Universit\"{a}t Heidelberg,
 Philosophenweg 16, 69120 Heidelberg, Germany}

\author{Fabian Rennecke}
\email[E-mail: ]{frennecke@bnl.gov}
\affiliation{Physics Department, Brookhaven National Laboratory,
  Upton, NY 11973, USA} 

\begin{abstract}
 Since the incident nuclei in heavy-ion collisions do not carry strangeness, the global net strangeness of the detected hadrons has to vanish. We investigate the impact of strangeness neutrality on the
  phase structure and thermodynamics of QCD at finite baryon and
  strangeness chemical potential. To this end, we study the
  low-energy sector of QCD within a Polyakov loop enhanced quark-meson
  effective theory with 2+1 dynamical quark flavors. Non-perturbative
  quantum, thermal, and density fluctuations are taken into account
  with the functional renormalization group. We show that the impact
  of strangeness neutrality on thermodynamic quantities such as the equation of state is sizable.
\end{abstract}

\maketitle

\section{Introduction}\label{sec:intro}

Ultrarelativistic heavy-ion collisions performed at RHIC and LHC aim
to explore the phase structure of quantum chromodynamics (QCD) at
finite temperature and density.  One of the key challenges is to
extract properties of the quark-gluon plasma (QGP) created in such
collisions from the hadronic final states that reach the detector. The
success of hadron resonance gas models (HRG), which are based on
thermal distributions of noninteracting hadrons, in describing various
aspects of the hadronization process might suggest that the system at
the time of freeze-out can be described by equilibrium thermodynamics
characterized by temperature and chemical potentials
\cite{BraunMunzinger:2003zd}.

Since the timescale of the weak interactions is much longer than the
equilibration time of the strongly interacting QGP, quark number
conservation of the strong interactions should hold from the initial
stage up to the freeze-out. So the strangeness and charge/isospin of
the incident nuclei determine the strangeness- and isospin chemical
potentials $\mu_S$ and $\mu_I$ at freeze-out. For instance, the
absence of strange quarks in nuclei implies strangeness neutrality,
i.e.~the net strangeness has to vanish. The baryon chemical potential
$\mu_B$, which is directly related to the baryon number at central
rapidity, additionally depends on the energy of the collision because
the rapidity distributions of net-baryon number show a distinctive
beam-energy dependence. In fact, this is the basis of current and
future beam-energy scan experiments aimed at exploring different
region of the QCD phase diagram \cite{Gazdzicki:995681, BESwp,
  Friman:2011zz, Kekelidze:2016hhw, GALATYUK201441, Sako:2014fha}.

To understand the properties of matter created in heavy-ion collisions
it is therefore indispensable to take these constraints into
account. Since quarks, mesons and baryons can carry finite strangeness
and isospin, the details of how these constraints are fulfilled depend
crucially on the state of QCD matter. Understanding this from a
theoretical point of view poses many challenges. The different phases
of QCD, including the dynamics of quarks, gluons and hadrons at
various temperatures and chemical potentials need to be
captured. Hence, purely hadronic effective models which are only valid
at the lowest energies and QCD perturbation theory only valid at very
high energies are only of limited use. Owing to the notorious sign
problem at finite $\mu_B$, lattice QCD simulations are restricted to
vanishing chemical potential. Nonetheless, tremendous progress has
been made in recent years in exploring the QCD equation of state at
finite $\mu_B$ on the lattice through, e.g.~the Taylor expansion of
the thermodynamical potential as a function of $\mu_B/T$ around
$\mu_B = 0$ \cite{Bazavov:2017dus} or the analytic continuation from
imaginary chemical potential \cite{Gunther:2016vcp}, among many other
approaches \cite{Muroya:2003qs, Aarts:2015tyj}. These techniques
allowed first studies of the freeze-out conditions of heavy-ion
collisions subject to the constraints on strangeness and isospin on
the lattice \cite{Bernard:2007nm, Bazavov:2012vg,
  Borsanyi:2013hza}. Since both methods rely on expansions in powers
of $\mu_B/T$, exploring regions of the phase diagram with
$\mu_B/T \gtrsim 1$ on the lattice might entail unknown and
potentially large systematic errors. For instance, at small beam
energies at RHIC the HRG predicts $\mu_B/T > 2$ at the freeze-out
\cite{Alba:2014eba}, so current experiments probe regions of the phase
diagram where state-of-the-art first principle methods might not be
fully reliable.

Functional continuum methods, such as the functional renormalization
group (FRG) and Dyson-Schwinger equations (DSE) do not suffer from the
sign problem, so the inclusion of finite chemical potential is
possible without the corresponding systematic errors. A lot of
progress has been made towards the study of QCD from first principles,
e.g.~\cite{Alkofer:2000wg, Fischer:2006ub,Maas:2011se,Bashir:2012fs,Fischer:2008uz,
Fischer:2014ata, Mitter:2014wpa, Cyrol:2017ewj} and references
therein. However, due to the necessity of truncating the effective
action of QCD, results at finite chemical potential from first
principles are currently only accessible with unknown and potentially
large systematic errors. Functional continuum methods are in some
sense complementary to the lattice, since the most common sources of
systematic errors on the lattice, such as finite-size effects, chiral
fermions and the sign problem, are not present in continuum methods
and, vice versa, the lattice does not have to rely on truncations of
the effective action.

Low-energy effective theories of QCD have proven time and again that
they can provide valuable insights on the QCD phase structure. Their
strength lies in the potential to identify physically relevant effects
that prevail also in the full theory. Prominent examples relevant for
the present work are Polyakov loop enhanced Nambu--Jona-Lasinio models
(PNJL), Polyakov loop enhanced quark-meson models (PQM) and (the
closely related) chiral matrix models. They can be constructed to
share the same global symmetries as QCD and exhibit similar or even
the same symmetry breaking patterns as the chiral transition of
QCD. Owing to the coupling to a non-vanishing gluon background field,
the deconfinement transition can also be captured in a statistical
manner \cite{Pisarski:2000eq}. In mean-field approximations, the phase structure and
thermodynamics of QCD have been studied in great detail with these
models, see e.g.~\cite{Meisinger:1995ih, Fukushima:2003fw,
  Ratti:2005jh, Sasaki:2006ww, Schaefer:2007pw, Fu:2007xc,
  Fukushima:2008wg, Sakai:2009dv,Pisarski:2016ixt} and
\cite{Fukushima:2017csk} for a recent review. In this context, the
constraint of strangeness neutrality has first been imposed in the
study of the phase structure in \cite{Fukushima:2009dx}. In compliance
with expectations from the HRG \cite{BraunMunzinger:2003zd} and the
lattice \cite{Bernard:2007nm}, it was demonstrated that a finite
strangeness chemical potential is necessary to ensure strangeness
neutrality at finite temperature and baryon chemical potential. This
is related to the intricate interplay of quark, meson and baryon
effects mentioned above.

Concerning strangeness and isospin dynamics, a major shortcoming of
mean-field studies is the lack of dynamics of the most relevant
degrees of freedom in the hadronic phase. Owing to their nature as
pseudo Goldstone bosons of spontaneous chiral symmetry breaking, these
are certainly pions and kaons regarding the effects related to isospin
and strangeness. It is therefore conceivable that their quantum
fluctuations have to be accounted for in order to accurately describe
the QCD medium as generated in heavy-ion collisions. A major challenge
is that hadronic fluctuations are in general of non-perturbative
nature. The FRG has been proven to be very useful here, since it
allows for the non-perturbative regularization and renormalization of
quantum fluctuations in low-energy models. For two flavors, the phase
structure and thermodynamics of (P)QM models have been studied
exhaustively with the FRG, e.g.~\cite{Berges:1997eu, Schaefer:2004en,
  Herbst:2010rf, Skokov:2010wb, Skokov:2010uh, Herbst:2013ail,
  Pawlowski:2014zaa, Fu:2015naa, Fu:2015amv, Fu:2016tey,
  Almasi:2017bhq}. These works carved out the crucial role of meson
fluctuations in the QCD equation of state. Finite isospin chemical
potential has been investigated in \cite{Kamikado:2012bt} within a QM
model. However, the constraint on isospin from heavy-ion collision has
not been considered in this work. Strangeness requires at least three
flavors. In this case, first studies of the phase structure with the
QM beyond mean-field have been carried out in \cite{Mitter:2013fxa,
  Rennecke:2016tkm, Resch:2017vjs} and the PQM at vanishing density
has been studied in \cite{Herbst:2013ufa}. Fluctuations in the strange
sector have been shown to be quantitatively and qualitatively relevant
for the phase structure of QCD in the former works. In the latter work
it has been demonstrated that lattice thermodynamics at vanishing
density can be reproduced by including fluctuations into the PQM model
with the FRG.

In this work we extend the previous works in two ways. The first is
the extension of \cite{Herbst:2013ufa} to finite baryon chemical
potential $\mu_B$ and the confrontation of the results on the equation
of state with most recent lattice results at finite $\mu_B$. Second,
and most importantly, we introduce a finite strange chemical potential
$\mu_S$ and derive the corresponding functional renormalization group
equations for the 2+1 flavor PQM. This allows us to impose the
strangeness neutrality condition on the equation of state in terms of
a $T$- and $\mu_B$-dependent $\mu_S$. As discussed above, this is a
property imprinted on the matter created in heavy-ion collision from
its initial conditions. For the first time, we are able to study the
influence of strangeness neutrality on the thermodynamics and phase
structure of QCD beyond mean-field directly at finite baryon chemical
chemical potential. Genuine finite density effects related to the
dynamics of strange hadrons are accessible this way. This is of
relevance for a general understanding of the properties of strongly
interacting matter as created in heavy-ion collisions.

This paper is organized as follows: In \Sec{sec:model} we introduce
the effective low-energy model used here, including a discussions of
the coupling of mesons to $\mu_S$ and the finite gluon background. The
functional renormalization group and the derivation of the
corresponding renormalization group equations is discussed in
\Sec{sec:flucts}. We present our results in \Sec{sec:res}. After the
discussion of the initial conditions for the solution of the RG
equations in \Sec{sec:ini}, we check the validity of our model by
comparing it to lattice results at vanishing and finite $\mu_B/T$ in
\Sec{sec:tdmu0}. In \Sec{sec:mus} we determine the strangeness
chemical potential neccesary to fulfill the strangeness neutrality
condition and discuss the role of quark, meson and baryon dynamics for
our results. In Sects.~\ref{sec:td} and \ref{sec:ps} we discuss the
influence of strangeness neutrality of the thermodynamics and the
phase structure of QCD. A summary and a brief outlook are given in
\Sec{sec:sum}. Details on the parametrization of the Polyakov loop
potential, thermodynamics at large $\mu_B$ and the initial conditions
are provided in the appendices.

\section{$N_f =2+1$ QCD at low
  energies}\label{sec:model}

Here we discuss the construction of a low-energy effective theory of
QCD that allows us to describe the main features of strangeness and
the phase structure on the same footing.

\subsection{Chemical potentials}\label{sec:chempots}

In QCD the numbers of each flavor are conserved separately. So in
general there is an independent chemical potential for each quark
flavor, e.g.\ \cite{Kogut:2001id}, 
\begin{align}\label{eq:muquark}
  \mu_u\,\bar u\gamma_0 u+ \mu_d\,\bar d\gamma_0 d+ \mu_s\,
\bar s\gamma_0 s\,.
\end{align}
The quark chemical potentials can be rewritten in terms of baryon-, strangeness- and isospin chemical potentials as follows
\begin{align}\label{eq:mus}
  \renewcommand\arraystretch{1.3}
 \mu=\begin{pmatrix} \mu_u 
\\ \mu_d \\ \mu_s \end{pmatrix} = 
  \begin{pmatrix} \frac{1}{3}\mu_B+\frac{1}{2}\mu_I 
    \\ \frac{1}{3}\mu_B-\frac{1}{2}\mu_I \\ \frac{1}{3}\mu_B-\mu_S \end{pmatrix}\,.
  \renewcommand\arraystretch{1.0}
\end{align}
We remark that on the lattice the quark chemical potentials are
typically written in terms of baryon-, strangeness- and charge
chemical potentials, leading to
\begin{align}\label{eq:mulat}
  \renewcommand\arraystretch{1.3}
 \mu= \begin{pmatrix} \frac{1}{3}\mu_{B,\rm lat}+\frac{2}{3}\mu_Q 
\\ \frac{1}{3}\mu_{B,\rm lat}-\frac{1}{3}\mu_Q \\ \frac{1}{3}\mu_{B,\rm lat}
-\frac{1}{3}\mu_Q-\mu_{S,\rm lat} \end{pmatrix}\,, 
  \renewcommand\arraystretch{1.0}
\end{align}
see e.g.\ \cite{Bernard:2007nm, Bazavov:2012vg,
  Borsanyi:2013hza}. Comparing the two definitions we infer that
$\mu_I=\mu_Q$ while $\mu_{B,\rm lat}=\mu_B-1/2\mu_Q$ and
$\mu_{S,\rm lat} = \mu_{S}-1/2 \mu_{Q}$. Note however, that
$\mu_B,\mu_{B,\rm lat}$ couple to the same operator $\bar q\gamma_0 q$
and baryon number fluctuations are either described with derivatives
w.r.t.\ $\mu_B$ or $\mu_{B,\rm lat}$. Moreover, for $\mu_I=\mu_Q=0$
the two definitions agree.

Hadrons carry charges associated to these chemical potentials, and
hence couple to the quark chemical potential $\mu_q$. This coupling
naturally emerges in the functional renormalization group approach
from an evolution of QCD from large momentum to low momentum scales
and the introduction of hadrons as effective low energy degrees of
freedom via dynamical hadronization
\cite{Gies:2001nw,Pawlowski:2005xe,Floerchinger:2009uf}, see 
\cite{Braun:2008pi, Braun:2014ata, Mitter:2014wpa, Rennecke:2015eba, Cyrol:2017ewj} for applications to
QCD. The coupling of the chemical potentials to hadrons then follows
directly from the Silver Blaze property of QCD \cite{Cohen:2003kd}. At
vanishing temperature, the chemical potential dependence of an
Euclidean $n$-point function of fields $\phi_i$ with associate
particle numbers $c_i$ is given by a simple shift of the external
frequency \cite{Marko:2013lxa,Khan:2015puu}
\begin{align}
p_{i,0} \,\rightarrow\, p_{i,0} + i c_i \mu_i\,.
\end{align}
Hence, one just needs to shift the frequencies of the kinetic
terms in the effective action according to the Silver Blaze
property. 

In the present low energy effective field theory setup it is simpler
to utilise a flavor symmetry argument, see
e.g.\ \cite{Kogut:2001id}. At its core this argument carries the Silver
blaze property of QCD discussed above, and it is straightforward to
check that both constructions yield the same result. Concentrating on
the mesons for the moment, we introduce the chemical potential as a
vector source. Then the chemical potential in \eq{eq:muquark} is
written as
\begin{align}\nonumber 
  C_\nu &\equiv \delta_{\nu 0}\, C\,,\\[1ex] 
  C &\equiv \text{diag}\Big(\frac{1}{3}\mu_B+
      \frac{1}{2}\mu_I,\frac{1}{3}\mu_B-\frac{1}{2}\mu_I,\frac{1}{3}\mu_B-\mu_S\Big)\,. 
\label{eq:C}\end{align}
Using this in the full quark part of the QCD Lagrangian we arrive at  
\begin{align}\label{eq:Lquark}
\mathcal{L}_q = \bar q \big( \gamma_\nu D_\nu+
\gamma_\nu C_\nu \big) q = \bar q \gamma_\nu \bar D_\nu\, q\,,
\end{align} 
with the modified covariant derivative $\bar D_\nu = D_\nu + C_\nu$
and $D_\mu=\partial_\mu - i\,g A_\mu$. This action is invariant under
an extended local $U(N_f)$ flavor symmetry if the vector source
$C_\nu$ transforms under local $U(N_f)$ transformations
$\mathcal{U}(x)$ as
\begin{align}\label{eq:UNf}
  C_\nu \,\rightarrow\, \mathcal{U}(x) C_\nu\, \mathcal{U}^\dagger(x) +
  \mathcal{U}(x) \partial_\nu\, \mathcal{U}^\dagger(x)\,, 
\end{align}
not to be confused with chiral flavor rotations.  Since the gauge part
of the modified covariant derivative is flavor-blind, gauge invariance
is trivially guaranteed. Scalar and pseudoscalar mesons are
represented as entries of a flavor matrix in the adjoint
representation of the flavor rotations defined in \eq{eq:UNf},
\begin{align}\label{eq:sigmadef}
\Sigma = T^a (\sigma_a+i \pi_a)\,. 
\end{align}
Here the generators are $T^0 = \mathds{1}/\sqrt{2 N_f}$ and
$T^{a\in\{1,\dots,N_f^2-1\}} \in SU(N_f)$. The meson sector inherits
the local flavor symmetry of the quark sector as described
above. Since the mesons transform in the adjoint representation, one
can immediately write down the corresponding covariant derivative,
\begin{align}\label{eq:mesoncov}
\bar D_\nu \Sigma = \partial_\nu \Sigma +[C_\nu,\Sigma]\,.
\end{align}
The chemical potential can be rewritten conveniently as
\begin{align}\label{eq:muBdiag}
\mu = \frac{1}{3}\mu_B\, \mathds{1}+ \text{diag}\Big(\frac{1}{2}\mu_I,
-\frac{1}{2}\mu_I,-\mu_S\Big)\,. 
\end{align}
With \eq{eq:muBdiag} and \eq{eq:mesoncov} it follows trivially that
the baryon chemical potential does not couple to the mesons, as it
should.  In turn, mesons are sensitive to strangeness and isospin. In
this work we assume light isospin symmetry and therefore set
$\mu_I = 0$.

\subsection{Low energy effective theory}\label{sec:let}

Here we discuss the low energy effective theory in terms of its
effective action. It has to captures the basic dynamics related to
strangeness at low energies. Dynamically most relevant are 
the kaons, since they are pseudo Goldstone bosons with strangeness
$\pm 1$. Chiral symmetry requires that if kaons are included in the
effective action, all other mesons in the lowest scalar and
pseudoscalar meson nonet have to be taken into account as well. This
can be understood intuitively by considering the quark-antiquark
scattering channels where the pseudoscalar kaons emerge as resonances,
\begin{align}\label{eq:kchan}\nonumber 
\mathcal{L}_K &\sim \big(\bar u \gamma_5 s\big)^2 + \big(\bar d \gamma_5 s\big)^2 + \big(\bar s \gamma_5 u\big)^2 + \big(\bar s \gamma_5 d\big)^2\\[1ex] \nonumber 
&\sim \big[\bar q \gamma_5 (T^4- i T^5) q\big]^2 + \big[\bar q \gamma_5 (T^6 + i T^7) q\big]^2 \\[1ex] 
&\quad+ \big[\bar q \gamma_5 (T^4+ i T^5) q\big]^2 + \big[\bar q 
\gamma_5 (T^6 - i T^7) q\big]^2\,,
\end{align}
where we choose the Gell-Mann matrices as $SU(N_f)$ generators.  In
terms of QCD flows for the effective action the four-fermi
interactions including their momentum-dependent couplings emerge from gluon exchange
diagrams. Note that the individual terms in \Eq{eq:kchan} can in
principle also have different couplings. However, it is
straightforward to show that this channel explicitly breaks
$U(N_f)_L\times U(N_f)_R$ chiral symmetry in any case. Since we are
also interested in the phase transition, the only allowed sources of
explicit chiral symmetry breaking are the current quark masses,
otherwise chiral symmetry restoration cannot be captured properly. The
four quark interaction channel that gives rise to a kaon resonance and
respects chiral symmetry is
\begin{align}\label{eq:4q}
  \mathcal L_K \subset \mathcal L_{4q} =
  \big( \bar q\, T^a q \big)^2 +  \big( \bar q\, i \gamma_5  T^a q \big)^2\,.
\end{align}
Bosonizing this channel via a standard Hubbard-Stratonovich
transformation \cite{Stratonovich, Hubbard:1959ub}, or
selfconsistently with dynamical hadronization, yields an effective action
containing the lowest scalar and pseudoscalar meson nonet as defined
in \Eq{eq:sigmadef}, including their coupling to quarks. Note that
\Eq{eq:4q} also contains the parity partners of the kaons, the kappas
(or $K_0^*$), as additional open-strange mesons. Chiral symmetry
dictates that we have to take them into account even though their mass
is above 1 GeV so they are dynamically irrelevant. Resonances with the
quantum numbers of pions, $\eta$, $\eta^\prime$, $f_0(980\!-\!1370)$ and
the critical modes of the chiral transition, the $\sigma$-mesons
($f_0(500)$), are also included in \Eq{eq:4q}. Note however, that the
identification of the heavy scalar meson is not entirely clear in our
case since we find a mass of about 1150 MeV, which is between the
known $f_0(980)$ and $f_0(1370)$ states. For more details on this construction see e.g.\ 
\cite{Rennecke:2016tkm}. Including these dynamical mesons, their
effective potential and coupling to quarks allows us to describe the
chiral phase transition.

Statistical confinement is included via a (temporal) gluon background
field $\bar A_\mu \equiv \bar A_0 \delta_{\mu0}$ and a
corresponding effective potential $U_\text{glue}(\bar A)$. This is
discussed in more detail in the next section. Putting all this
together gives rise to a Polyakov loop enhanced quark-meson (PQM)
model with 2+1 dynamical quark flavors at finite baryon and
strangeness chemical potential.  It is an approximation for the full
effective action of low energy QCD valid below momentum scales
$k\lesssim \Lambda$ with the ultraviolet cutoff scale
$\Lambda \!\sim\! 1$ GeV. By definition $\Lambda$ is the scale below
which gluons decouple from the matter sector of QCD, and hence
constituent quarks and hadrons in a gluon background field provide a
good description of QCD. We will elaborate on this further in
\Sec{sec:flucts}. 

In the current work we use the following approximation to the full
scale-dependent Euclidean effective action of the 2+1 flavor PQM
model,
\begin{align}\label{eq:ea}
  \Gamma_k &= \int_x\Bigl\{
             \bar q \bigl(\gamma_\nu D_\nu+\gamma_\nu C_\nu\bigr)q + h\,
             \bar q\,\Sigma_5 q\\ \nonumber
           &\quad+\text{tr}\bigl(\bar D_\nu\Sigma\!\cdot\!\bar D_\nu
             \Sigma^\dagger\bigr)+\widetilde U_k(\Sigma , \bar A) + U_\text{glue}(\bar A)
             \Bigr\}\,.
\end{align}
In \eq{eq:ea} quantum, thermal and density fluctuations of modes with
Euclidean momenta $\Lambda \geq |p| \geq k$ have been integrated
out. The gauge covariant derivative is
$D_\nu = \partial_\nu - i g \bar A_\nu$ and
$\Sigma_5 = T^a (\sigma_a+i \gamma_5\pi_a)$. The effective meson
potential $\widetilde U_k(\Sigma, \bar A)$ consist of a fully
$U(N_f)_L\times U(N_f)_R$ symmetric part plus pieces that explicitly
break subgroups of the full chiral symmetry group,
\begin{align}\label{eq:pot}
  \widetilde U_k(\Sigma, \bar A) = U_k(\rho_1,\tilde\rho_2, \bar A) - j_l \sigma_l 
  - j_s \sigma_s -c_A \xi\,.
\end{align}
$U_k$ is the chirally symmetric part of the meson potential. $j_l$ and
$j_s$ are explicit chiral symmetry breaking sources that account for
the finite current quark masses of the light and the strange
quarks. As before, we assume light isospin symmetry.  The 't Hooft
determinant $\xi = \det(\Sigma)+\det(\Sigma^\dagger)$ effectively
incorporates the anomalous breaking of $U(1)_A$
\cite{Kobayashi:1970ji, Hooft:1976up, Hooft:1976fv}. For
simplicity, we restrict ourselves to two out of a total of $N_f$
chiral invariants,
\begin{align}
  \rho_1 = \text{tr}\, \Sigma\Sigma^\dagger\,,\qquad \tilde \rho_2 
  = \text{tr}\Big(\Sigma\Sigma^\dagger -\frac{1}{2}\rho_2 \mathds{1}\Big)^2\,.
\end{align}
With the total effective potential $V_k = \widetilde U_k + U_\text{glue}$ and the solution $\bar \Phi_k(T,\mu_B,\mu_S)$ of the equations of motion,
\begin{align}
\frac{\partial V_k(\Phi)}{\partial \Phi}\bigg|_{\bar \Phi_k} = 0\,,
\end{align}
where $\Phi = (\Sigma,\bar A)$, the $k$-dependent thermodynamic potential is given by
\begin{align}
\Omega_k = V_k(\bar \Phi_k)\,.
\end{align}
It can be used to define the cumulants of baryon number and
strangeness,
\begin{align}\label{eq:chis}
  \chi_{ij}^{BS} = -T^{i+j-4}\, \frac{\partial^{i+j} 
  \Omega_0(T,\mu_B,\mu_S)}{\partial\mu_B^i\partial\mu_S^j}\,.
\end{align}
Net baryon number and strangeness are given by the first cumulants,
and their densities are obtained by dividing out the spatial volume
$\mathcal{V}$,
\begin{align}
  \nonumber 
  n_B &=  \0{\langle N_B -N_{\bar B} \rangle}{\mathcal{V}} = 
        \chi_{10}^{BS}\, T^3 \,,\\[1ex]
  n_S &= \0{\langle N_{\bar S} - N_S \rangle}{\mathcal{V}} =
        \chi_{01}^{BS}\, T^3\,,
\label{eq:ns}
\end{align}
Note that strange antiquarks are defined to have
$\langle S\rangle =n_S \mathcal{V}=1$. In the presence of a large strange
chemical potential it might be necessary to take the difference
between the light and strange sectors into account also in the symmetric
part of the effective potential. This can be achieved by first
redefining the generators such that they decompose into purely strange
and non-strange parts,
\begin{align}
  \renewcommand\arraystretch{1.2}
 \begin{pmatrix} \widetilde T^0 \\  \widetilde T^8 \end{pmatrix} 
 = \frac{1}{\sqrt{3}}
 \begin{pmatrix} \sqrt{2} & 1 \\ 1 & -\sqrt{2} \end{pmatrix}
  \begin{pmatrix} T^0 \\  T^8 \end{pmatrix}\,,
\label{eq:brot}
\renewcommand\arraystretch{1.0}
\end{align}
while keeping 
\begin{align}
\widetilde T^{a\in\{1,\dots,7\}} &= T^{a\in\{1,\dots,7\}}\,. 
\end{align}
\Eq{eq:brot} is the rotation from the singlet-octet to the
light-strange basis of $U(N_f)$. The respective fields are 
\begin{align}\nonumber 
  \Sigma^{(L)} &= \widetilde T^{a\in\{ 0,1,2,3\}} (\sigma_a 
                 + i \pi_a)\,,\\[2ex]
  \Sigma^{(S)} &=\widetilde T^{a\in\{4,5,6,7,8 \}} (\sigma_a 
                 + i \pi_a)\,.
\label{eq:lsinv}
\end{align}
$\widetilde T^{a\in\{ 0,1,2,3\}}$ are generators of $U(2)$, but
embedded in $U(3)$. Since $\Sigma^{(S)}$ contains all generators with
non-vanishing off-diagonal entries in the third column and/or row, it
contains the open strange mesons, i.e., those with strangeness
$S= \pm 1$. With this, the new invariants can straightforwardly be
constructed. Note that there are no mixed invariants since
$\text{tr}\, T^a T^b = \delta^{ab}/2$. But for now, we will not do
this and work with the fully symmetric potential $U_k$. This is a good
approximation as long as the strange chemical potential is not too
large. For instance, At $T=0$ and $\mu_B = 0$ one expects kaon
condensation if $\mu_S \gtrsim m_K$. In this case, one would certainly
have to construct the effective action based on the fields in
\Eq{eq:lsinv}. But as we discuss below, we are only interested in
strange chemical potentials $\mu_S \lesssim 200$ MeV where \Eq{eq:pot}
is expected to be a good approximation.

\subsection{Gluonic background}\label{sec:gluepot}

The Euclidean action of $SU(N_c)$ Yang-Mills theory at finite
temperature $T$ is invariant under `twisted' gauge transformations
$\mathcal{U}$ which obey for $\beta = 1/T$
\begin{align}
\mathcal{U} (x_0 + \beta, \vec{x}) = z_n\, \mathcal{U} (x_0, \vec{x})\,,
\end{align}
where $z_n$ is an element of the center of the gauge group, i.e.~$z_n = \mathds{1} e^{i 2 \pi n/N_c}$ for $SU(N_c)$. The Polyakov loop \cite{Polyakov:1978vu},
\begin{align}\label{eq:poldef}
L(\vec{x}) = \frac{1}{N_c} \tr\, \mathcal{P} e^{i g \int_0^\beta\! dx_0 A_0(x_0, \vec{x})}\,,
\end{align}
where $\mathcal{P}$ is the path ordering and the trace is in the
fundamental representation, is invariant under gauge transformations but not under center transformations, $L\rightarrow z_n L$. The expectation
value of the Polyakov loop is related to the free energy $ F_{q\bar q}$ of a 
quark-antiquark pair at infinite distance \cite{McLerran:1981pb},  
\begin{align}\label{eq:Lex} 
\langle L \rangle \sim e^{-\012 \beta F_{q\bar q}}\,. 
\end{align}
In \eq{eq:Lex} we have used declustering and
$\langle \bar L \rangle=\langle L \rangle$. Confinement implies that
it takes an infinite energy to remove the antiquark from the system,
and hence $F_{q\bar q}$ has to be infinity.  Accordingly
$\langle L \rangle = 0$. In the deconfined phase the free energy of an
isolated quark is finite and thus $\langle L \rangle \neq 0$. Hence,
the Polyakov loop serves as an order parameter for the deconfinement
transition in the static limit, which can be associated to the
breaking/restoration of center symmetry.

In the spirit of the present mean-field theory for gluons the Polyakov
loop is taken into account by a temporal gluonic background
$\bar A_\mu=\delta_{\mu0} \bar A_0$, as already mentioned before. As
the effective action is invariant under background gauge
transformations, the (constant) background gauge field can be rotated
into the Cartan subalgebra, to wit,
\begin{align}\label{eq:a0}
\0{g}{2\pi T} \bar A_0 &= \0{g}{2\pi T} \left( \bar A_0^{(3)} t^3 + \bar A_0^{(8)} t^8\right) \\ \nonumber
&=\0{\varphi_3}{2}  \left(\begin{array}{crc} 1& 0 & 0\\ 0& -1 & 0\\ 0 & 0 & 0\end{array}\right) 
+ \0{\varphi_8}{ 2\sqrt{3}} \left(\begin{array}{ccc} 1& 0 & 0\\ 0& 1 & 0\\ 0 & 0 & -2\end{array}\right) \,,
\end{align}
where we defined
\begin{align}
\varphi_i = \0{g \bar A_0^{(i)}}{2 \pi T}\,,\qquad i=3,8\,, 
\end{align}
for the eigenvalues of the temporal gauge field. 
Inserting this
into \Eq{eq:poldef}, the integral and trace become trivial and the
Polyakov loop and antiloop are:
\begin{align}
L &= \frac{1}{3} e^{i\0{\pi}{\sqrt{3}}  \varphi_8} \Big( e^{-i\sqrt{3}\pi  \varphi_8}+
2\cos( \pi \varphi_3) \Big)\,,\\[1ex]
\bar L &= \frac{1}{3} e^{-i \0{\pi}{\sqrt{3}}  \varphi_8} \Big( e^{i\sqrt{3}\pi  
\varphi_8}+2\cos( \pi\varphi_3) \Big)\,.
\end{align}
Since we are working in a field theoretical approach with a gauge
field $A_\mu$ we should use $L[\langle A_0 \rangle] = L[\bar A_0]$, instead of
$\langle L[A_0]\rangle$ as computed on the lattice \cite{Braun:2007bx,Marhauser:2008fz}. The former
variable shows a more rapid transition from the confined to the
deconfined phase, and is saturated by unity for temperatures
$T\gtrsim 1.25\, T_c$. The difference is accounted for with a trivial,
but temperature-dependent normalisation factor, for more details see
\cite{Herbst:2015ona}. In the present work we use a mean field
approximation for the glue dynamics leading to 
$L[\langle A_0 \rangle]= \langle L[A_0 ]\rangle$. This approximation 
will be lifted in future work. 
 
Note also 
that our effective action \eq{eq:ea} is manifestly gauge invariant
since the gluon background field only appears in the covariant
derivative of the quarks and the gauge invariant Polyakov loops, which
are the variables of the gluon effective potential as discussed below.

The idea underlying the above formulation has been proven to be very
successful in Matrix- or Polyakov-loop models, where the simple
representation of the gluon field in \eq{eq:a0} leads to particularly
simple expression of $L$, while still being able to capture main
features of confinement, see e.g.\ \cite{Fukushima:2017csk} and
references therein. By now this has been also worked out for full QCD
\cite{Braun:2007bx,Braun:2009gm,Herbst:2015ona}, which provides a
natural embedding of the current model into QCD as a QCD-assisted effective
field theory, e.g. \cite{Pawlowski:2014aha}. 

At finite chemical potential another intricacy has to be taken care
of: since quarks and antiquarks manifest themselves in the
effective action with terms
\begin{align}
L\, e^{-\mu_q/T}\,, \quad \text{and} \quad \bar L\, e^{\mu_q/T}\,,
\end{align}
in the fermion occupation numbers, they have to be real valued in
order to give a well-defined equation of state. Here, we defined the quark chemical potential $\mu_q = \mu_B/3$. Furthermore, at finite
chemical potential they are also unequal. Hence, while one can assume
without loss of generality that $\varphi_8 = 0$ at $\mu = 0$, it has to be
non-zero and imaginary at finite $\mu$,
\begin{align}
\bar \varphi_8= - i \varphi_8 \,, \quad \bar \varphi_8\in\mathds{R}\,.
\end{align}
The loops then are
\begin{align}
\begin{split}
  L &= \frac{1}{3} e^{-\0{\pi }{\sqrt{3}}\bar \varphi_8}
  \Big( e^{\sqrt{3}\pi \bar \varphi_8}+2\cos(\pi \varphi_3) \Big)\,,\\
  \bar L &= \frac{1}{3} e^{\0{\pi }{\sqrt{3}}\bar \varphi_8} \Big( e^{-\sqrt{3}\pi \bar \varphi_8}
  +2\cos( \pi \varphi_3) \Big)\,.
\end{split}
\end{align}
This was pointed out, e.g.,\ in \cite{Dumitru:2005ng,Reinosa:2015oua,
  Pisarski:2016ixt, Maelger:2017amh}. In practice, the transition from a QM to a PQM
model can be achieved by a simple replacement of the quark
distribution function, $n_F \rightarrow N_F$, in many cases. The
reason is that the $\bar A_0$ eigenvalues enter the computation as a $SU(N_c)$-valued
imaginary shift of the chemical potential, cf.~\Eq{eq:Lquark}. Hence,
in any finite-temperature loop computation where the chemical
potential only enters through the Fermi-Dirac distribution, the
non-trivial color trace (i.e.~the sum over the eigenvalues) simply
results in a modified distribution function,
\begin{align}\label{eq:nf}
&N_F(E_q,\mu_q; L,\bar L)\\ \nonumber
&\quad= \frac{1 + 2 \bar L e^{(E_q-\mu_q)/T}+L e^{2(E_q-\mu_q)/T}}{1 + 3 \bar L e^{(E_q-\mu_q)/T}+3 L e^{2(E_q-\mu_q)/T} + e^{3(E_q-\mu_q)/T}}\,,
\end{align}
where $E_q$ is the quark quasiparticle energy,
$E_q = \sqrt{k^2+m_q^2}$, where $k$ is the modulus of the spatial
momentum. But note that we pointed out in \cite{Fu:2016tey} that this
simple replacement is not always correct. The modified distribution
function has a very useful qualitative interpretation: in the confined
phase with $L \approx 0$ one has
$N_F \approx 1/\big\{\!\exp[3 (E_q-\mu_q)/T] +1 \big\}$, which is the
distribution function for a $qqq$-state, a baryon. See
\cite{Fu:2015naa} for a more careful discussion of this behavior. In
the deconfined phase $N_F$ is identical to the distribution of a
single quark. The terms $\exp[2 (E_q-\mu_q)/T]$ in \Eq{eq:nf} can be
interpreted as intermediate diquark states. So the coupling of the
gluon background field $\bar A_0$ to the quarks leads to a smooth
interpolation between baryons in the hadronic phase and quarks in the
QGP. Even though the effective action in \Eq{eq:ea} only has mesons as
explicit hadronic content, we can still account for baryon
dynamics. Including both a baryon- and a strange chemical potential
allows us to capture the effects of strange and nonstrange baryons
separately.

To be able to capture the deconfinement phase transition, an effective
gluon potential is necessary. The strategy for Polyakov-loop enhanced
models is to use a phenomenological parametrization of the effective
potential of the pure gauge theory at finite temperature in terms of
Polyakov loops. In this work we use the parametrization introduced in
\cite{Lo:2013hla} with
$U_\text{glue}(\bar A) = U_\text{glue}(L,\bar L)$ given by
\begin{align}\label{eq:plpot}
\begin{split}
\frac{U_\text{glue}(L,\bar L)}{T^4}  &= -\frac{1}{2} a(T) \bar L L + b(T) \ln\big[M_H(L,\bar L)\big]\\
&\quad+\frac{1}{2} c(T) (L^3+\bar L^3) + d(T) (\bar L L)^2\,,
\end{split}
\end{align}
where $M_H$ is the $SU(3)$ Haar measure in terms of the Polyakov loops,
\begin{align}
M_H(L,\bar L) = 1-6 \bar L L + 4(L^3+\bar L^3)-3(\bar L L)^2\,.
\end{align}
The advantage of this parametrization is that it reproduces the
pressure and the Polyakov loop susceptibilities of $SU(3)$ Yang-Mills
theory. The relevance of an accurate description of Polyakov loop
susceptibilities in particular for the cumulants of particle number
distributions has been discussed in \cite{Fu:2015naa,
  Fukushima:2017csk} and explicitly demonstrated in
\cite{Fu:2016tey}. The explicit choice for the parameters $a$, $b$ and
$c$ is discussed in \App{app:poloop}. There, we also discuss how the
chemical potential dependence of the Polyakov loop potential is
modelled.

By relying on a parametrization of the gauge potential based only on
Yang-Mills theory, we make sure that all effects related to matter
fluctuations, i.e.~the unquenching, are included dynamically within
our model through the coupling of $\bar A_0$ to the quarks. Since this
is not put in by hand here, it adds to the predictive power of the
model.

\section{Fluctuations}\label{sec:flucts}

It has
been shown that even for zero chemical potentials at the very least pion
fluctuations are required to get reasonably accurate results
for the QCD equation of state \cite{Herbst:2013ufa}. We argued that
for strangeness dynamics kaons are the most relevant degrees of
freedom at small and moderate chemical potentials as they are the
lightest strange particles in the hadronic sector. So without kaon
fluctuations crucial effects related to finite $\mu_S$ would certainly
be missed. To account for meson fluctuations we use the functional
renormalization group. It is a semi-analytical method providing a
non-perturbative regularization and renormalization scheme for the
resummation of an infinite class of Feynman diagrams. For reviews of
the FRG we refer the reader to \cite{Berges:2000ew, Pawlowski:2005xe,
  Gies:2006wv, Schaefer:2006sr, Delamotte:2007pf, Rosten:2010vm,
  Braun:2011pp}.

\subsection{The Functional Renormalization Group}\label{sec:frg}

The FRG realizes Wilson's renormalization group idea of successively
integrating out quantum fluctuations from large to small energy
scales. The starting point is the microscopic action
$\Gamma_{k=\Lambda}$ at some large initial momentum scale $\Lambda$ in
the UV. By lowering the RG scale $k$, quantum fluctuations are
successively integrated out until one arrives at the full macroscopic
quantum effective action $\Gamma \equiv \Gamma_{k=0}$ at
$k=0$. Ideally, one starts in the perturbative regime where the
initial effective action $\Gamma_{k=\Lambda}$ is related to the
well-known microscopic action of QCD. As already discussed before, in
the present low-energy approach we choose $\Lambda$ at a scale where
we assume that gluon degrees of freedom are already integrated
out. Hence, $\Lambda$ is directly linked to the Yang-Mills mass gap
with $\Lambda \lesssim 1\,\text{GeV}$. In Landau gauge QCD the
Yang-Mills mass gap is reflected in the gapping of the gluon
propagator which leads to an effective suppression of gluonic diagrams 
in a functional approach such as the FRG, see the reviews 
\cite{Alkofer:2000wg,Fischer:2006ub,Bashir:2012fs,Maas:2011se} and references therein. 

The FRG formulates the RG in terms of a functional differential
equation for the evolution of the scale dependent effective action
$\Gamma_k$, the Wetterich equation \cite{Wetterich:1992yh,Ellwanger:1993mw,Morris:1993qb}.
In the present case, with dynamical quarks and mesons in a gluon
background, the flow equation reads
\begin{align}\nonumber 
  \partial_t \Gamma_k &= \frac{1}{2}\sum_{i=1}^{2N_f^2} \Tr \,
                        \left(\!G_{\phi_i \phi_i,k} \!\cdot \partial_t R_k^{\phi_i}\right)\\[1ex]
                      &\quad - 2\Tr \left(G_{l\bar l ,k} \!\cdot \partial_t R_k^l\right)- \Tr\, 
                        \left( G_{s\bar s,k} \!\cdot \partial_t R_k^s\right)\,,
\label{eq:fleq}\end{align}
where $\partial_t = k\frac{ d}{ d k}$ denotes the logarithmic scale
derivative. The trace runs over all discrete and continuous indices,
i.e. color, spinor and the loop momenta and/or frequencies
respectively. The sum in the first line is over all $2 N_f^2$ scalar
and pseudoscalar mesons in \Eq{eq:sigmadef}. The generalized meson and
quark propagators are given by matrix elements in field space,   
\begin{align}\label{eq:propdef}
\begin{split}
G_{\Phi_i\Phi_j,k}[\Phi] &= \left[\0{1}{\Gamma^{(2)}_k[\Phi] + R_k}
\right]_{\Phi_i\Phi_j}\!\!(p,-p)\,,
\end{split}
\end{align}
with the generalized field $\Phi=(\phi, q, \bar q, \bar A_0)$, $R_k$ is the matrix of regulators 
$R_k^{\phi_i}, R^l_k, R_k^s$ being diagonal for the mesons and symplectic for the quarks, and 
$\Gamma^{(2)}_k=\delta^2\Gamma_k/\delta\Phi^2$. Since we assume
isospin symmetry we define the light quark as $l \equiv u = d$ and the
quark field becomes $q = (l,l,s)$. The scale-dependent IR regulators
$R_k^{\Phi_i}$ can be understood as momentum-dependent masses that
suppress the infrared modes of the field $\Phi_i$. In addition, the
terms $\partial_t R_k^{\Phi_i}$ in \Eq{eq:fleq} also ensure
UV-regularity. Their definitions and a more explicit form of the flow
equation will be discussed in the next section. We use the local
potential approximation (LPA) here, which means that only the
symmetric part of the meson effective potential, $U_k$, is running in
\Eq{eq:ea}. For a study of effects beyond LPA in the QM at finite
temperature and density we refer to \cite{Pawlowski:2014zaa,
  Rennecke:2016tkm}. While effects beyond LPA are certainly relevant,
at least the qualitative features of the relevant physics for the
present purposes are captured by the running of the effective
potential.

The FRG is a method to integrate out quantum fluctuations in Euclidean
spacetime in terms of one-particle irreducible (1PI)
diagrams. Consequently, the dynamics is driven by quantum fields
propagating as internal lines of 1PI Feynman diagrams with Euclidean
momenta. All interactions are governed by off-shell fields. This
implies a very simple hierarchy for dynamically relevant
contributions: the lighter the degree of freedom, the more relevant it is. This
means in particular that the contribution of particles with masses
$m \gtrsim \Lambda$ to, for instance, the equation of state, is
negligible. Within this fluctuation-driven approach one therefore
expects that kaons and $s$-quarks coupled to $\bar A_0$ are sufficient
to capture the relevant strangeness effects at small to moderate
chemical potentials in the same way that the dynamics of pions and
quarks coupled to $\bar A_0$ already give almost quantitative results
for the equation of state at vanishing chemical potentials,
cf.~\cite{Herbst:2013ufa}. This is in contrast to purely statistical
approaches without quantum fluctuations, such as the HRG \cite{BraunMunzinger:2003zd}, where the lack of dynamics and
interactions has to be compensated by taking into account all possible
hadrons and their excited states. While being very successful in the
description of particle properties at the freeze-out, the QCD phase
transition and features of the QGP are not accessible in such
approaches.

\subsection{Flow of the effective potential}\label{sec:fleqs}

Here, we briefly discuss the RG flow equations of our model. For
$\mu_S = 0$ this has been discussed in \cite{Herbst:2013ufa,
  Mitter:2013fxa, Rennecke:2016tkm, Resch:2017vjs}. We therefore focus
on the manifestly new contributions to the flow equation here. As
discussed in Sec.~\ref{sec:chempots}, the non-vanishing strange
chemical potential also couples to the open strange mesons. In our
case these are the four scalar kappa-mesons and the four
pseudoscalar kaons. Induced by the covariant derivative $\bar D_\nu$
in \Eq{eq:mesoncov}, this leads to a shift of the frequency in the
kinetic terms of these particles. All other mesons are unaffected by
finite $\mu_S$. Their contributions to the flow of the effective
potential is therefore identical to the ones in, e.g.,
\cite{Rennecke:2016tkm}. We will therefore only outline the changes
for the open strange mesons. For definiteness, we pick out the
contribution of the charged kaons, $K^\pm$. Within the present approximation, the regulated propagator
defined in \Eq{eq:propdef} is:
\begin{align}\nonumber 
&G_{K^+K^-,k}(p_0,\vec{p}\,;\mu_S)\\[1ex]
&= \frac{1}{(p_0-i \mu_S)^2 +\vec{p}^{\,2}\big(1+r_B(
\vec{p}^{\,2})\big)+m_{K,k}^2}\,,
\end{align}
where the delta distribution for momentum conservation is
omitted. Note that finite $\mu_S$ leads to a
linear frequency term in the propagator. We choose to regulate only
the spatial momenta with a regulator of the form
$R_k^\phi = \vec{p}^{\,2}\, r_B(\vec{p}\,)$. Nonetheless,
both UV and IR regularity for arbitrary frequencies is still
guaranteed. We use the flat or Litim regulator with the shape function
$r_B(\vec{p}^{\,2}) = (k^2/\vec{p}^{\,2}-1) \Theta(k^2-\vec{p}^{\,2})$
\cite{Litim:2000ci, Litim:2001up}. For the antiparticle propagator,
only the sign of $\mu_S$ changes,
\begin{align}
G_{K^-K^+,k}(p_0,\vec{p}\,;\mu_S) = G_{K^+K^-,k}(p_0,\vec{p}\,;-\mu_S)\,.
\end{align}
Inserting this into the flow equation \eq{eq:fleq}, we find
\begin{align}\label{eq:oscontr}
\nonumber
&\frac{1}{2} \text{STr}\, G_{K^+K^-,k}\, \partial_t {R}_{k}^{K}\\[1ex] \nonumber
&= \frac{1}{2} T \sum_{n\in \mathds{Z}} \int\!\!\frac{d^3p}{(2\pi)^3} G_{K^+K^-,k}(\omega_n,\vec{p},\mu_S)\, \partial_t {R}_{k}^{K}(\vec{p}\,)\\[1ex] \nonumber
&= \frac{k^4}{12 \pi^2} \frac{k}{E_K} 
\Big[ 1\!+ n_B(E_K \!-\! \mu_S) + n_B(E_K \!+\! \mu_S) \Big]\\[1ex]
&\equiv \frac{k^4}{4 \pi^2}\, \bar l_0^{\,(K)}(\mu_S)  \,,
\end{align}
where $\omega_n \!=\! 2\pi n T$ is the bosonic Matsubara frequency,
$n_B(E) \!=\! [\exp(E/T) -1]^{-1}$ is the Bose-Einstein distribution,
$E_K \!=\! \sqrt{k^2+m_{K,k}^2}$ is kaon energy. In this form the thermal particle,
antiparticle as well as the vacuum contribution of open strange mesons
are manifest. Since this expression is symmetric under exchange of
particles and antiparticles ($\mu_S\rightarrow -\mu_S$), it also holds
for the $K^-K^+$-contributions as well as for $K^0$ and $\bar
K^0$. For the contribution of the $\kappa$'s, only the quasiparticle
energy has to be replaced,
$E_K \rightarrow E_\kappa$.

The flow of the effective potential in terms of the physical fields is given by 
\begin{align}\label{eq:uflow}
\nonumber
 &\partial_{t} U_k(\rho_1,\tilde\rho_2) = \\ \nonumber
& \frac{k^4}{4\pi^2} \bigg\{ \bar l_0^{\,(f_0)}(0) + 3 \bar l_0^{\,(a_0)}(0)+4\bar l_0^{\,(\kappa)}(\mu_S)+ \bar l_0^{\,(\sigma)}(0)\\ \nonumber
& +  \bar l_0^{\,(\eta)}(0) +3 \bar l_0^{\,(\pi)}(0) +4 \bar l_0^{\,(K)}(\mu_S) + \bar l_0^{\,(\eta^\prime)}(0) \\ 
&- 4N_c  \Big[ 2 \bar l_0^{(l)}(\mu_q) + \bar l_0^{(s)}(\mu_q-\mu_S) \Big] \bigg\}\,,
\end{align}
with the quark threshold function
\begin{align}
\begin{split}
 &\bar l_0^{(q)}(\mu)\\
 &= \frac{k}{3 E_q} \Big[ 1\!- N_F(E_q,\mu;L,\bar L) + \bar N_F(E_q,\mu;L,\bar L) \Big]\,,
 \end{split}
\end{align}
and the Fermi-Dirac distribution in presence of a non-vanishing $A_0$
background $N_F$ \eq{eq:nf}. The antiquark distribution
function is given by $\bar N_F(E_l,\mu;L,\bar L) = N_F(E_l,-\mu;\bar L,L)$. \Eq{eq:uflow} is identical to the one used in \cite{Rennecke:2016tkm}, except that $\mu_S$ now enters the threshold
functions of the open strange mesons through the distribution
function in \Eq{eq:oscontr}.

\subsection{Flow of the particle numbers}\label{sec:Sfleq}

The computation of the cumulants of particle number distributions
require derivatives of the thermodynamic potential with respect to the
chemical potential, cf.~\Eq{eq:chis}. While it is simple to perform
these derivatives numerically, many points in $\mu_{B,S}$ are required
to ensure numerical accuracy and for higher cumulants this is
practically not feasible. One alternative is to use algorithmic
derivation techniques, see e.g.~\cite{Wagner:2009pm}. The other
alternative is given by solving the flow equations for the cumulants
directly. For first discussions in this direction we refer to
\cite{Fu:2015naa, Almasi:2017bhq}. In both cases, the accuracy of a
cumulant of arbitrary order is given by the accuracy of the
differential equation solver that is used and numerical derivatives on
the data are obsolete. We will not give an exhaustive discussion here and
restrict ourselves to the cases directly relevant for the present
work.

It is straightforward to derive flow equations for the cumulants. For
the first cumulants, i.e.~the particle numbers, this is particularly
simple due to
\begin{align}
 \frac{d \Omega_k}{d\mu} = 
  \frac{\partial \Omega_k}{\partial\mu}+ \frac{\partial 
  \Omega_k}{\partial\Phi}\frac{\partial\Phi}{\partial\mu} = 
  \frac{\partial \Omega_k}{\partial\mu}\,,
\end{align}
where $\Phi$ contains all meson and quark fields as well as the
Polyakov loop and antiloop. In the last step, the equations of motion
were used. Hence, only the explicit dependence of the effective
potential on $\mu$ is relevant here. Within the LPA we use in the
present work, only the effective potential is running and, under the
assumption that one can interchange the RG scale derivative and the
$\mu$-derivative, a simple flow equation for the strangeness number
density $n_S$ is obtained, 
\begin{align}
\begin{split}
\partial_t n_{S,k} &= -\frac{k^4}{\pi^2}\Big[ \partial_{\mu_S} \bar l_0^{\,(\kappa)}(\mu_S) + \partial_{\mu_S} \bar l_0^{\,(K)}(\mu_S)\\
&\quad - N_c \partial_{\mu_S} \bar l_0^{\,(s)}(\mu_q-\mu_S) \Big]\,.
\end{split}
\end{align}
As discussed above and in \App{app:poloop}, the Polyakov loop
potential $U_\text{glue}$ also carries an explicit $\mu_S$
dependence. Since $U_\text{glue}$ does not run, we can store its
contribution into the initial condition for convenience. If the
initial action would be $\mu_S$-independent, the initial strangeness would
then be trivially given by
$n_{S,\Lambda} = - \partial_{\mu_S} \Omega_\Lambda =
- \partial_{\mu_S} U_\text{glue}$. However, as we discuss in the next
section, there is an important in-medium correction to the initial
potential, $\Delta\Gamma_\Lambda$, so we provide the explicit equation
for the initial strangeness number in the next section.

Since the mesons do not carry baryon number, the flow equation for the
corresponding density is just given by the fermion contribution,
\begin{align}
\begin{split}
\partial_t n_{B,k} &= \frac{N_c k^4}{\pi^2} \partial_{\mu_B}\Big[2 
\bar l_0^{\,(l)}(\mu_q) + \bar l_0^{\,(s)}(\mu_q-\mu_S) \Big]\,.
\end{split}
\end{align}
Again we store the $k$-independent gluon contribution in the initial
conditions. This will be discussed in the next section.

\section{Results}\label{sec:res}

\subsection{Initial Conditions}\label{sec:ini}

The scale set by temperatures above the critical temperature $T_c$ exceeds the cutoff scale
$\Lambda$ of the effective model, $2\pi T \gtrsim \Lambda$. In order
to describe thermodynamic quantities above $T_c$, we therefore need
initial conditions that depend on the temperature and, since we are
interested in finite chemical potential effects as well, also on
$\mu$. These initial conditions are governed by integrating out
fluctuations from scales $\bar\Lambda \gg 2\pi T$ down to
$\Lambda$. Hence, we want to correct our vacuum initial conditions for
in-medium effects at the initial scale, for a recent detailed discussion 
see \cite{Braun:2018svj}. This is achieved by
integrating the initial vacuum effective action from $\Lambda$ to
$\bar \Lambda$ and subsequently integrating the in-medium effective
action down to $\Lambda$ again \cite{Braun:2003ii},
\begin{align}\label{eq:medinit}\nonumber
&\Delta\Gamma_\Lambda(T,\mu_q,\mu_S)\\ 
&\quad= \int_\Lambda^{\bar\Lambda} 
\frac{dk}{k}\Big[\partial_t\Gamma_k(0,0,0)-\partial_t\Gamma_k(T,\mu_q,\mu_S)\Big]\,.
\end{align}
As long as the scale set by the medium parameters is smaller than
$\Lambda$, $\Delta\Gamma_\Lambda(T,\mu_q,\mu_S)$ vanishes because the
in-medium flow and the vacuum flow are identical for $k \geq
\Lambda$. Since quark fluctuations certainly dominate over meson
fluctuations for $\Lambda \gtrsim 900\,\text{MeV}$, we can approximate
the flows in \Eq{eq:medinit} by the purely fermionic ones, to wit,
\begin{align}\label{eq:medinitexpl}
  \nonumber
  &\Delta\Gamma_\Lambda(T,\mu_q,\mu_S)\\ \nonumber
  &\quad= -\int_\Lambda^{\infty} \!dk\, \frac{N_c k^4}{3 \pi^2}\bigg
    \{\frac{2}{E_l}\Big[N_F(E_l,\mu_q;L,\bar L)\\ \nonumber
  &\qquad+\bar N_F(E_l,\mu_q;L,\bar L)\Big]+\frac{1}{E_s}
    \Big[N_F(E_s,\mu_q - \mu_S;L,\bar L)\\
  &\qquad+\bar N_F(E_s,\mu_q - \mu_S;L,\bar L)\Big]\bigg\} \,.
\end{align}
We set
$\bar\Lambda \!\rightarrow\! \infty$ since the thermal contribution to
the quark flow is UV regular.

\begin{table}[t]
\begin{center}
\begin{tabular}{ c | c  }
\hline \hline
parameter & value\\
\hline
$\Lambda$ & $0.9 \,\text{GeV}$ \\
$\lambda_{10,\Lambda}$ & $(0.830 \,\text{GeV})^2$ \\
$\lambda_{20,\Lambda}$ & $10$ \\
$\lambda_{01,\Lambda}$ & $54$ \\
$h$ & 6.5 \\
$j_l$ & $(0.121 \,\text{GeV})^3$ \\
$j_s$ & $(0.336 \,\text{GeV})^3$ \\
$c_A$ & $4.808 \,\text{GeV}$ \\
\hline
$b_0$ & $1.6$ \\
$\alpha_t$ & $0.47$ \\
\hline \hline
\end{tabular}
\end{center}
\caption{Parameters for the initial effective action and the Polyakov loop potential. They are chosen such that we find in the vacuum at $k=0$ for the pion mass $m_\pi = 138$ MeV, for the kaon mass $m_K = 495$ MeV, for the $\sigma$-meson mass $m_\sigma = 463$ MeV, for the sum $m_\eta^2+m_{\eta^\prime}^2 = 1.218$ GeV$^2$, for the light current quark mass $m_l = 302$ MeV and for the decay constants $f_\pi = 93$ MeV and $f_K = 113$ MeV. The last two parameters belong to the Polyakov loop potential and are fixed by the pressure of 2+1 flavor lattice QCD at vanishing chemical potentials, see \App{app:poloop}.}
\label{tab:ini}
\end{table}

\begin{figure*}[t]
\centering
\includegraphics[width=.32\textwidth]{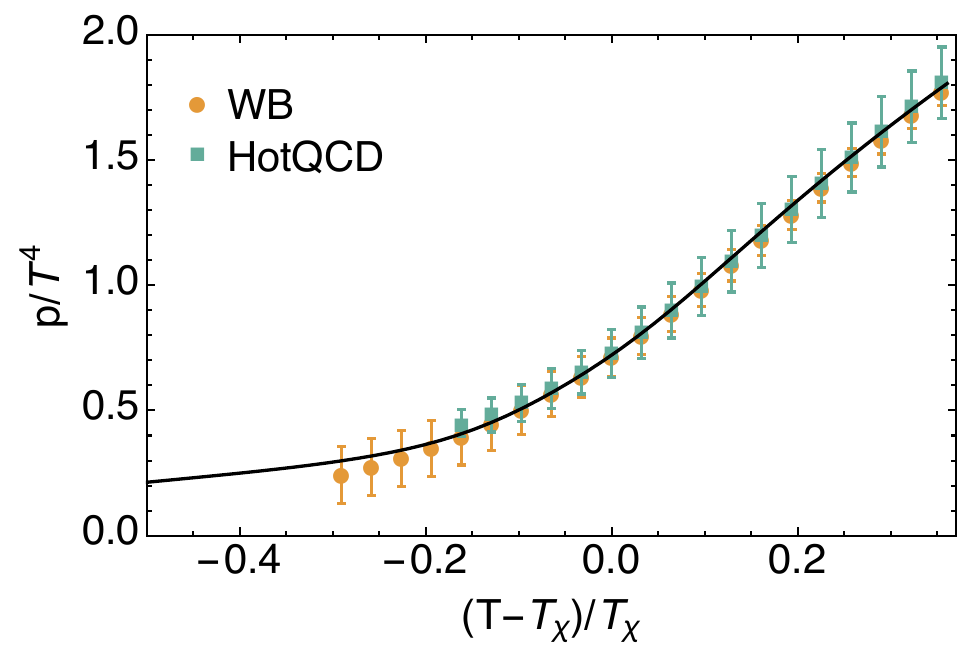}
\includegraphics[width=.31\textwidth]{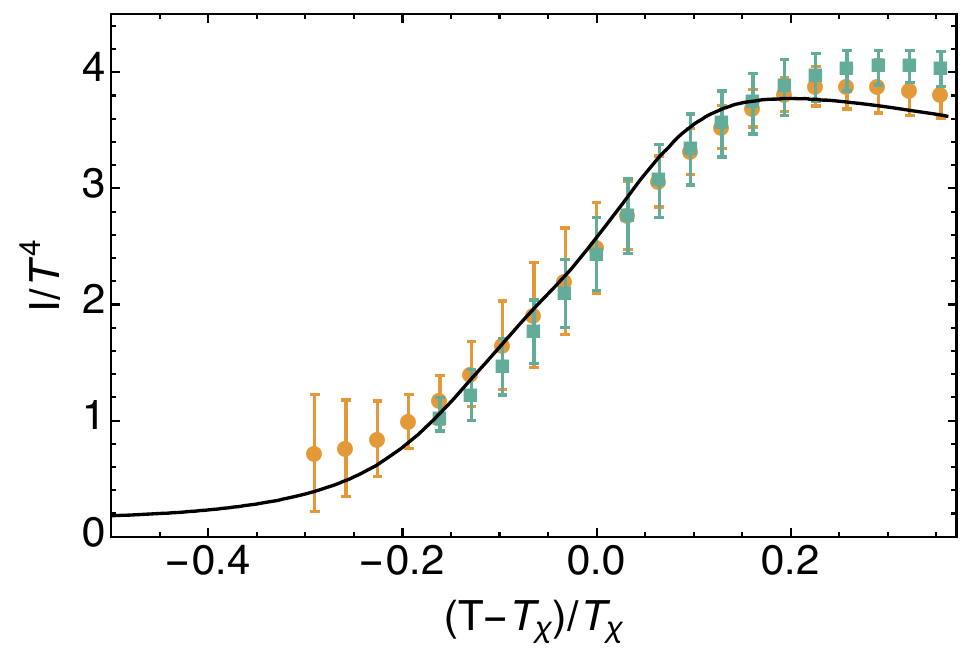}
\includegraphics[width=.32\textwidth]{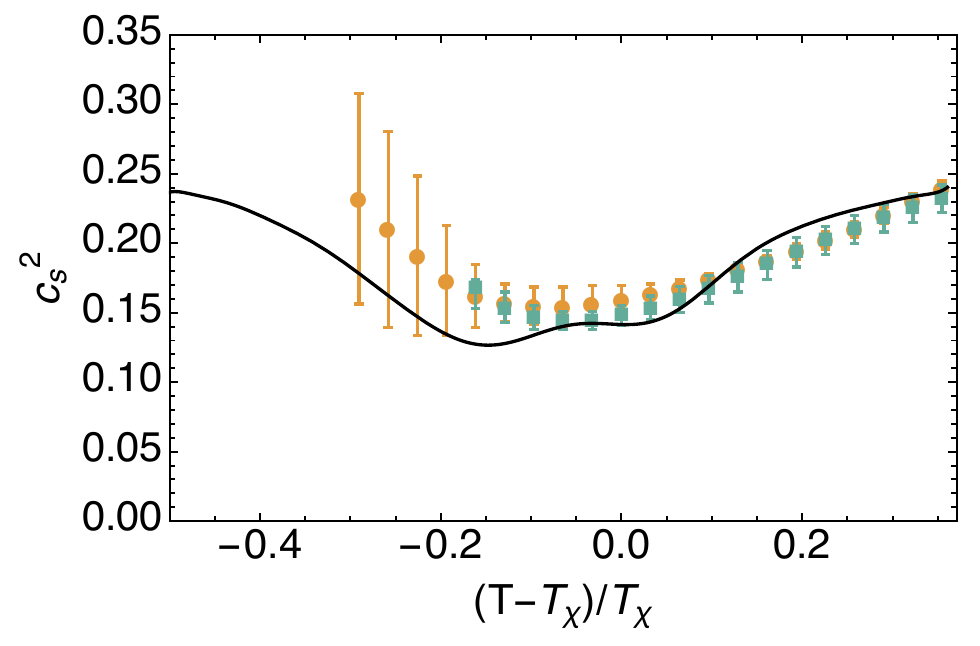}
\caption{The pressure $p$, the trace anomaly $I$ and the speed of
  sound squared $c_s^2$ at $\mu_B = \mu_S = 0$ in comparison to
  lattice results. The temperature has been rescaled to
  $t \equiv (T-T_\chi)/T_\chi$ du to different pseudocritical
  temperatures in our model and on the lattice. The HotQCD
  collaboration data is from \cite{Bazavov:2014pvz} and the
  Wuppertal-Budapest collaboration (WB) data from
  \cite{Borsanyi:2013bia}.}
\label{fig:thermolat}
\end{figure*}

It is important to note that $\Delta\Gamma_\Lambda$ not only depends
on the medium parameters but also on the field expectation values. The
dependence on the gluon background field in the current mean field
approximation for the glue dynamics enters through the Polyakov loops
$L,\bar L=\langle L[A_0]\rangle, \langle \bar L[A_0]\rangle$, and the
meson field expectation values through the quark masses. Since the
Polyakov loop expectation values approach their deconfined value only
for $T \gtrsim 4T_c$, cf.~\cite{Bazavov:2009zn}, non-trivial values
for $L,\bar L$ have to be taken into account in
\Eq{eq:medinitexpl}. Note that this may change when going beyond the mean
field approximation for the glue dynamics. As discussed before,
$L[\langle A_0\rangle]$ approaches unity far more rapidly \cite{Herbst:2015ona}.

Furthermore, if the meson part of the effective
potential is computed away from its stationary point, the relevant
quark masses are those given by $m_l= h \sigma_l/2$ and
$m_s = h \sigma_s/\sqrt{2}$, where $\sigma_l$ and $\sigma_s$ are the
meson background fields which, in general, do not have to coincide
with their vacuum expectation values as long as one is still able to
reliably solve the corresponding equation of motion for the mesons
(e.g.~by sampling the potential on a grid of field configurations as
in \cite{Mitter:2013fxa, Resch:2017vjs} or by using the fixed
background Taylor expansion as in \cite{Pawlowski:2014zaa,
  Rennecke:2016tkm}). With all the background- and medium-dependencies
spelled out explicitly, the initial potential is
\begin{align}
\begin{split}
&\Omega_\Lambda(\sigma_l,\sigma_s,L,\bar L; T,\mu_B,\mu_S)\\
&\quad= \widetilde U_{\Lambda}(\sigma_l,\sigma_s) 
+\Delta\Gamma_\Lambda(\sigma_l,\sigma_s,L,\bar L; T,\mu_B,\mu_S)\\
&\qquad+ U_\text{glue}(L,\bar L; T,\mu_B,\mu_S)\,,
\end{split}
\end{align}
where we added the $U_\text{glue}$ for convenience. Since it does not depend on the RG scale $k$, it is irrelevant whether we add it to the initial or to the final potential. Since it also carries no dependence on the meson fields, it only contributes to the pressure and leaves the initial meson $n$-point functions unaffected. The initial meson potential is
\begin{align}\label{eq:inittu}
\begin{split}
\widetilde U_\Lambda(\sigma_l,\sigma_s) &= U_\Lambda(\rho_1,\tilde\rho_2)-j_l\sigma_l-j_s\sigma_s-c_A\frac{\sigma_l^2 \sigma_s}{2\sqrt{2}}\\
&= \lambda_{10,\Lambda}\,\rho_1+\frac{1}{2}\lambda_{20,\Lambda}\,\rho_1^2+\lambda_{01,\Lambda}\,\tilde\rho_2\\
&\quad -j_l\sigma_l-j_s\sigma_s-c_A\frac{\sigma_l^2 \sigma_s}{2\sqrt{2}}\,.
\end{split}
\end{align}
It is sufficient to take only relevant and marginal operators into account at the initial scale since meson fluctuations are small at high energies and irrelevant operators are dimensionally suppressed in addition. Note that irrelevant operators are generated by the RG flow at smaller scales and are quantitatively and qualitatively relevant \cite{Pawlowski:2014zaa}. Our initial values are listed in \Tab{tab:ini}. The last two parameters are free parameters of the Polyakov loop potential and are discussed in \App{app:poloop}. In general, these initial parameters have uncertainties related to the uncertainties in the masses and decay constants we use to fix them. However, these uncertainties are irrelevant within the scope of the present work and are therefore neglected.

The total contribution to the initial conditions for mesonic $n$-point functions can be expanded as:
\begin{align}
\begin{split}
&U_\Lambda(\rho_1,\tilde\rho_2) +\Delta\Gamma_\Lambda(\sigma_l,\sigma_s,L,\bar L; T,\mu_B,\mu_S)\\
&\quad= \sum_{n,m=0}^N\frac{\omega_{nm,\Lambda}}{n!m!}(\rho_1-\kappa_1)^n(\tilde\rho_2-\kappa_2)^m\,,
\end{split}
\end{align}
and as a consequence of the discussion above the expansion coefficients are
\begin{align}\label{eq:medinipot}
\omega_{nm,\Lambda} = \lambda_{nm,\Lambda} + \frac{\partial^{n+m} \Delta\Gamma_\Lambda}{\partial\rho_1^n\partial\tilde\rho_2^m}\bigg|_{\kappa_1,\kappa_2}\,.
\end{align}
Following \Eq{eq:inittu} only the renormalizable initial parameters of the chirally symmetric part of the effective potential, $\lambda_{10,\Lambda},\,\lambda_{20,\Lambda},\,\lambda_{01,\Lambda}$, are nonzero. However, due to the meson background field dependence of $\Delta\Gamma_\Lambda$, these and higher order initial couplings receive medium- and gluon background dependent corrections. As the explicit symmetry breaking parameters $j_l$, $j_s$ and $c_A$ do not run within the present approximation, they are unaffected. We discuss viable simplifications of these complicated initial conditions in \App{app:fdi}.

The initial conditions for flows of the particle numbers are also affected by $\Delta\Gamma_\Lambda$. As discussed in the previous section, we store the contribution of the glue potential in the initial conditions for convenience. Thus, we find for the the strangeness and baryon number densities:
\begin{align}
\begin{split}
n_{S,\Lambda} &= -\partial_{\mu_S}\Delta\Gamma_\Lambda - \partial_{\mu_S}U_\text{glue}\,,\\
n_{B,\Lambda} &= -\partial_{\mu_B}\Delta\Gamma_\Lambda - \partial_{\mu_B}U_\text{glue}\,.
\end{split}
\end{align}
The system of flow equation is solved by using the fixed background Taylor expansion developed in \cite{Pawlowski:2014zaa, Rennecke:2016tkm}.

\subsection{Comparison to lattice gauge theory}\label{sec:tdmu0}

\begin{figure*}[t]
\centering
\includegraphics[width=.32\textwidth]{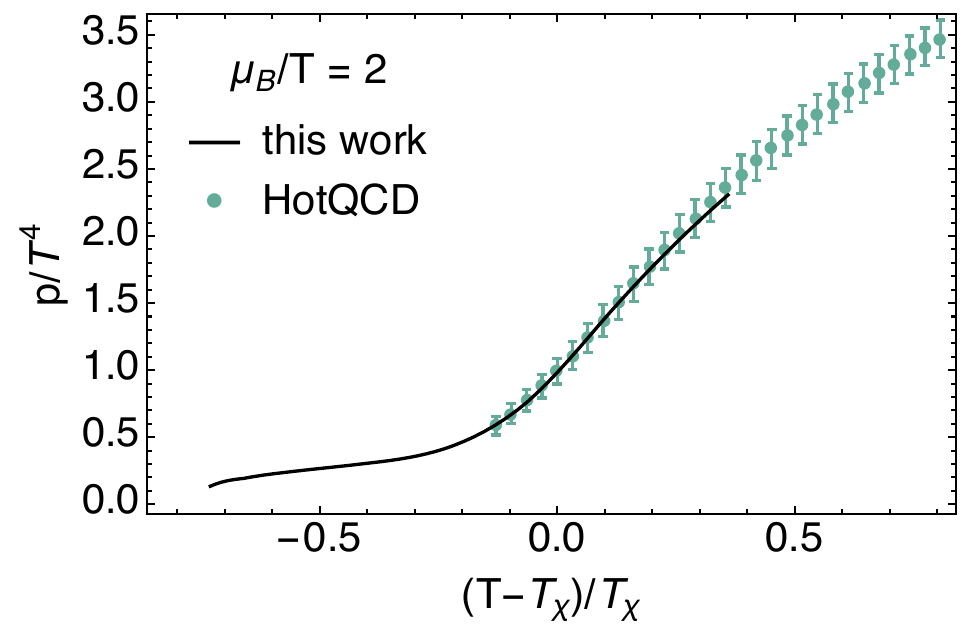}
\includegraphics[width=.32\textwidth]{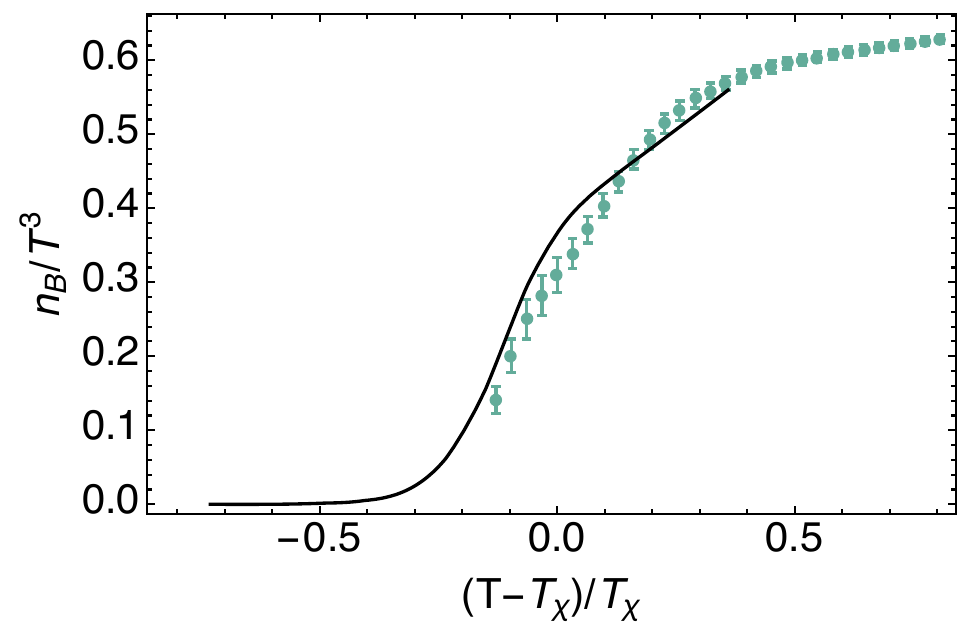}
\includegraphics[width=.329\textwidth]{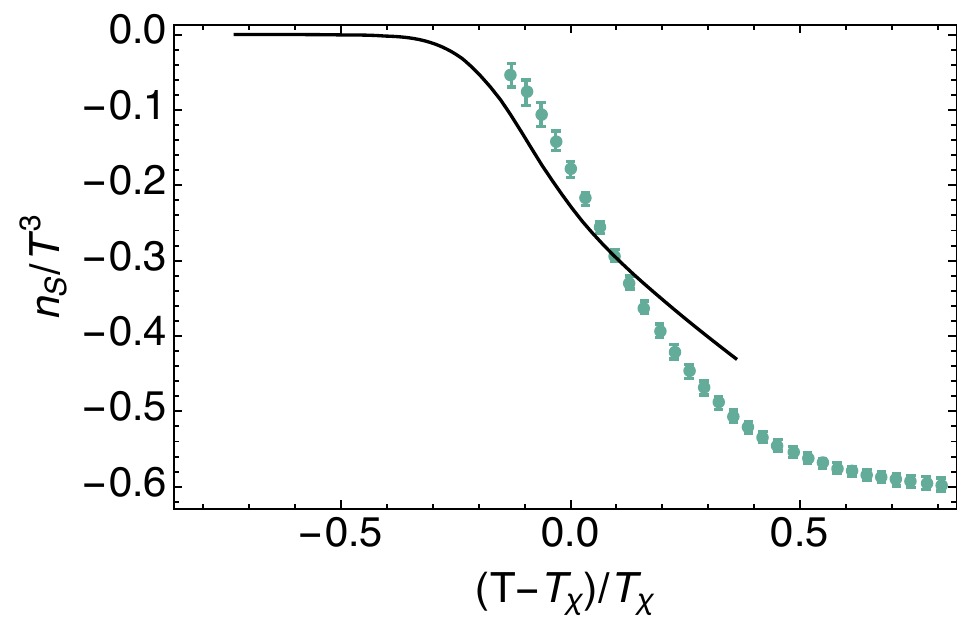}
\caption{The pressure $p$, the baryon number density $n_B$ and the
  strangeness density $n_S$ at $\mu_B/T = 2$ and $\mu_S = 0$ in
  comparison to lattice results as a function of the rescaled
  temperature $t$. The lattice data is taken from
  \cite{Bazavov:2017dus}.}
\label{fig:thermolatmu}
\end{figure*}

To demonstrate the validity of our model at vanishing chemical potentials, we compare our results on thermodynamic quantities to the results of lattice gauge theory. Within our model, the pseudocritical temperature of the chiral transition, which we define as the location of the inflection point of the subtracted chiral condensate,
\begin{align}\label{eq:subcond}
\Delta_{LS} = \frac{\big(\sigma_L - \frac{j_L}{j_S} \sigma_S\big)\big|_{T}}{\big(\sigma_L - \frac{j_L}{j_S} \sigma_S\big)\big|_{T=0}}\,,
\end{align}
is $T_\chi = 176.5$ MeV. This is roughly 15\% larger than the pseudocritical temperature found on the lattice \cite{Borsanyi:2010bp} so the absolute scale in our computation differs from the lattice. We therefore use relative temperature scales $t = (T-T_\chi)/T_\chi$ for our comparison. This allows us to compare the overall shapes of the functions which are sensitive to the relevant dynamics. The pressure, $p$, entropy density, $s$, energy density, $\epsilon$, trace anomaly, $I$, and the speed of sound squared, $\tilde{c}_s^{\,2}$, are defined as follows,
\begin{align}\label{eq:thermodefs}
\begin{split}
p &= -\Omega_0\,,\\
s &= \frac{\partial p}{\partial T}\,,\\
\epsilon &= - p + T s + \mu_B n_B + \mu_S n_S\,, \\
I &= \epsilon - 3 p\,,\\
\tilde{c}_s^{\,2} &=\frac{s(T,\mu_B,\mu_S)}{\partial \epsilon(T,\mu_B,\mu_S)/ \partial T} \,.
\end{split}
\end{align}
We note that, strictly speaking, $\tilde{c}_s^{\,2}$ is identical to the (hydrodynamic) speed of sound only at vanishing density, since the latter is defined at fixed entropy and particle numbers. We use $\tilde{c}_s^{\,2}$ since we want to compare different thermodynamic quantities for fixed values of the baryon chemical potential and different strangeness chemical potentials.

Our results on the pressure, the interaction measure and the speed of
sound squared in comparison to the lattice are shown in
\Fig{fig:thermolat}. We single out the trace anomaly and the speed of
sound since they are sensitive to the particle number densities and to
temperature derivatives of the pressure. We find excellent agreement
with lattice results for the pressure and the trace anomaly and good
agreement for the speed of sound. But note that the former has been
used to fix the free parameters of the Polyakov loop potential,
cf.~the last two parameters in Tab.~\ref{tab:ini}. The speed of sound
squared is thermodynamically highly nontrivial since it involves two
$T$ derivatives of the pressure. Furthermore, since it is a ratio of
two extensive thermodynamic quantities (the entropy and the heat
capacity) that grow with the number of degrees of freedom, this
effect, which dominates in particular the behavior of the pressure at
large $T$, is cancelled to some extent. The two minima of $c_s^2$ in
our computation are due to the fact that we find quite different
pseudocritical temperatures of deconfinement, $T_d$, and the chiral
transition, with $T_d \approx 155$ MeV if defined as the inflection
point of $L(T)$. The first minimum $c_s^2$ then corresponds to the
deconfinement transition and the second to the chiral transition.

To check the validity of our simple model also at finite $\mu_B$ we
compare it to lattice results obtained from a Taylor expansion of the
thermodynamic potential for various $\mu_B/T$ at $\mu_S = 0$
\cite{Bazavov:2017dus}. \Fig{fig:thermolatmu} shows the results for
$\mu_B/T = 2$. We note that the comparison does not change
qualitatively for other ratios. Only the temperature is rescaled for
comparison but we assumed that the ratio $\mu_B/T$ is the same for our
calculation and the lattice. This means that for instance at
$t\approx 0.35$ we have $\mu_B = 480$ MeV in our calculation and
$\mu_B = 420$ MeV in the lattice results. We have chosen the chiral
transition temperatures $T_\chi$ for $\mu_B=0$ for the definition of
$t$. With this, the pressure shows perfect agreement with the lattice
even at finite $\mu_B$. The same is true for the entropy density not
shown here.

Most sensitive to the finite-$\mu_B$ effects are certainly the
particle numbers, since they are only generated by finite chemical
potentials in the first place. We therefore also compare our results
on $n_B$ and $n_S$ to the lattice results in
\Fig{fig:thermolatmu}. The baryon number density agrees very well with
the lattice results at $\mu_B/T = 2$. In contrast to the lattice, we
see a larger bump in the vicinity of the phase transition. Note that
the bump appears in the lattice data only at the highest order in the
expansion of the thermodynamic potential presently available, which is
$\mu_B^6$ \cite{Bazavov:2017dus}. The error on the lattice data stems
from the determination of the expansion coefficients for a given
order. The systematic error, e.g., from missing higher-order
corrections of the expansion, is unknown. So it is possible that the bump
becomes more pronounced in the lattice data at higher orders of the
expansion. The strangeness density drops less steep with $t$ in our
results, but the overall agreement is still good. We want to emphasize
that the difference between $n_B$ and $-n_S$ in our computation stems
solely from the fluctuations of open strange mesons at $\mu_S = 0$. So
within a mean-field study of the (P)QM/(P)NJL models the physical
difference between $n_B$ and $-n_S$ in the hadronic phase at vanishing
$\mu_S$ cannot be captured.

The discrepancy between our results and
the lattice results for $n_S$ at larger $t$ could be a hint that
strange baryon dynamics are not captured quantitatively in the PQM
model. As discussed in \Sec{sec:gluepot}, they enter indirectly
through the coupling to the gluon background field. This appears to
work very well for $n_B$, on the other hand, indicating that nucleon
effects are described well. The three-quark states that contribute
through the modified fermion distribution function in \Eq{eq:nf}
always contain the same quark flavor, so while $lll$-states such as
the nucleons or $sss$-states such as the $\Omega$ are effectively
taken into account, the dynamically most relevant strange baryons, the
$lls$-states $\Lambda$ and $\Sigma$, but also $lss$-states such as the
$\Xi$ might not be captured accurately here. This could, rather heuristically,
explain the very good agreement of $n_B$ and the small deviations of
$n_S$.

\subsection{$\mu_S$ at strangeness neutrality}\label{sec:mus}

\begin{figure}[t]
\centering
\includegraphics[width=.49\textwidth]{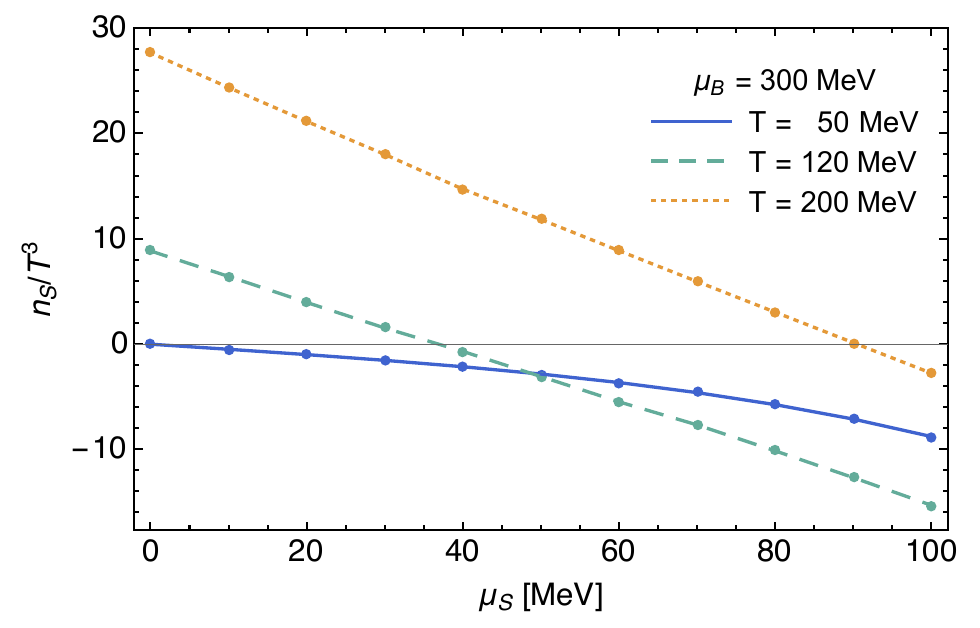}
\caption{Strangeness density as a function of $\mu_S$ at $\mu_B = 300$
  MeV for various temperatures.}
\label{fig:nsofmus}
\end{figure}

We computed the strangeness density
$n_S(T,\mu_B,\mu_S)$ for $T \in \{20,\dots,250\}$ MeV and
$\mu_B, 3\mu_S \in \{0,\dots,675\}$ MeV. We note that the low-energy effective theory is only valid up to moderate chemical potentials so we refrain from exploring the region beyond $675$ MeV. This is discussed in detail in \App{app:tdlamu}. An example of $n_S$ as a
function of $\mu_S$ for fixed $\mu_B$ and different $T$ is given in
\Fig{fig:nsofmus}. It is interesting to observe that $n_S$ is a linear
function of $\mu_S$ at larger temperatures. The larger $\mu_B$, the
smaller the temperature where this linear behavior emerges. Given that
$n_S/T^3 = \chi_{01}^{BS}$, we conclude that higher strangeness
cumulants $\chi_{0n}^{BS}$ for $n \geq 3$ are highly suppressed at
moderate to large temperatures.

\begin{figure}[t]
\centering
\includegraphics[width=.49\textwidth]{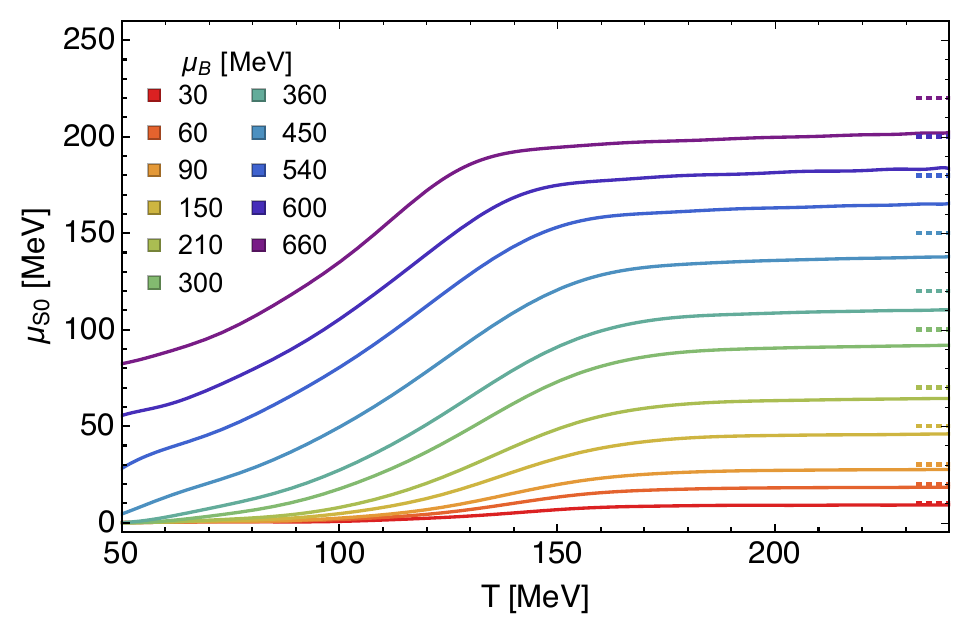}
\caption{The strange chemical potential as a function of temperature
  at strangeness neutrality for various baryon chemical potentials. The asymptotic values for free quarks
  are indicated by the dotted lines at the right edge of the
  plot. $\mu_B$ is increasing from bottom to top.}
\label{fig:musovert}
\end{figure}

The zero crossing of $n_S$ gives the value of $\mu_S$ that enforces
strangeness neutrality for given $T$ and $\mu_B$. Put differently,
$n_S = 0$ implicitly defines the function
\begin{align}
\mu_{S0}(T,\mu_B) = \mu_S(T,\mu_B)\Big|_{n_S = 0}\,.
\end{align}
In \Fig{fig:musovert} we show our results of $\mu_{S0}$ as a function
of $T$ for various $\mu_B$ at strangeness neutrality. We see that it
is always a monotonously increasing function of $T$ for the baryon
chemical potentials considered here. At large temperatures we find
$\mu_{S0} \approx \mu_B/3$, as indicated by the dashed lines at the
right edge of the figure. Furthermore, at small temperatures,
$T \approx 50$ MeV, $\mu_{S0}$ becomes nonzero only for
$\mu_B \gtrsim 400$ MeV. For $\mu_B = 0$ $\mu_{S0}$ is zero for all
$T$. Qualitatively, these observations can be understood as follows:
Since the baryon chemical potential couples to all quark flavors
equally, cf.~\Eq{eq:mus}, increasing $\mu_B$ will also increase the
number of strange quarks over antistrange quarks in the system. The
strange chemical potential, on the other hand, favors antistrange over
strange quarks and can therefore be tuned to compensate the
strangeness generated by $\mu_B$. Obviously, if $\mu_B$ is zero, than
$\mu_S$ also has to be zero to ensure strangeness neutrality. In the
hadronic phase at small $\mu_B$ essentially all strangeness is carried
by open strange mesons, in particular kaons and antikaons, since they
can always be excited in the thermal medium. At small temperatures the
Fermi surface of the baryons is very sharp while their Fermi energy is
large, so at small $\mu_B$ and small $T$ essentially no baryons are
excited. The thermally excited mesons will always have as much open
strange as open antistrange in the case of isospin symmetry
($\mu_I = 0$) for $\mu_S = 0$. Hence, $\mu_{S0} \approx 0$ at small T
and $\mu_B$. At large enough $\mu_B$ baryons can be excited and a
finite $\mu_S$ becomes necessary to ensure strangeness neutrality. The
corresponding strangeness will either be carried mostly by kaons (and
$\kappa$) or by baryons, depending on $\mu_B$.

With increasing temperature the Fermi surface of baryons becomes
increasingly diffused, facilitating the excitation of baryons. Hence,
$\mu_S$ has to increase accordingly with temperature to maintain
$n_S= 0$. This explains why $\mu_{S0}$ is monotonously increasing with
temperature. In the vicinity of the phase transition, mesons and
baryons start to dissolve into quarks. In the deconfined phase at
large $T$ the quarks are only weakly interacting and hence flavor is
decorrelated. In this case, there is an exact relation between baryon
number and strangeness that directly follows from the coupling of
$\mu_B$ and $\mu_S$ to the quarks in \Eq{eq:Lquark}. This implies
$\mu_{S0} = \mu_B/3$ in the deconfined phase. Since we find that the
Polyakov loops are still smaller than one even at $T = 250$ MeV
(characterizing the so called semi-QGP phase), complete deconfinement
is not reached for highest temperatures in \Fig{fig:musovert}, which
explains the the deviation of $\mu_{S0}$ from its asymptotic value.

Finally, we want to compare or findings to the predictions of a purely
fermionic system. In \cite{Fukushima:2009dx} an intriguing relation
between the Polyakov loops and the strangeness chemical potential at
strangeness neutrality has been derived,
\begin{align}\label{eq:fuku}
\mu_{S0}(T,\mu_B) \approx \frac{\mu_B}{3} - \frac{T}{2} 
\ln\!\bigg[\frac{\bar L(T,\mu_B)}{ L (T,\mu_B)}\bigg]\,.
\end{align}
The independence of the Polyakov loops on $\mu_S$ was assumed
here. This equation can be derived from the quark contribution to the
flow of the effective potential in \Eq{eq:uflow}. It provides a good
measure for the effect of the quarks coupled to the gluon background
field on strangeness neutrality. For the mean-field PNJL model studied
in \cite{Fukushima:2009dx} it has be shown to be be about 3\%
accurate. Potential deviations from this relation could be induced by
a strong $\mu_S$-dependence of $\bar L / L$ and, most importantly,
fluctuations of open strange mesons.

\begin{figure}[t]
\centering
\includegraphics[width=.49\textwidth]{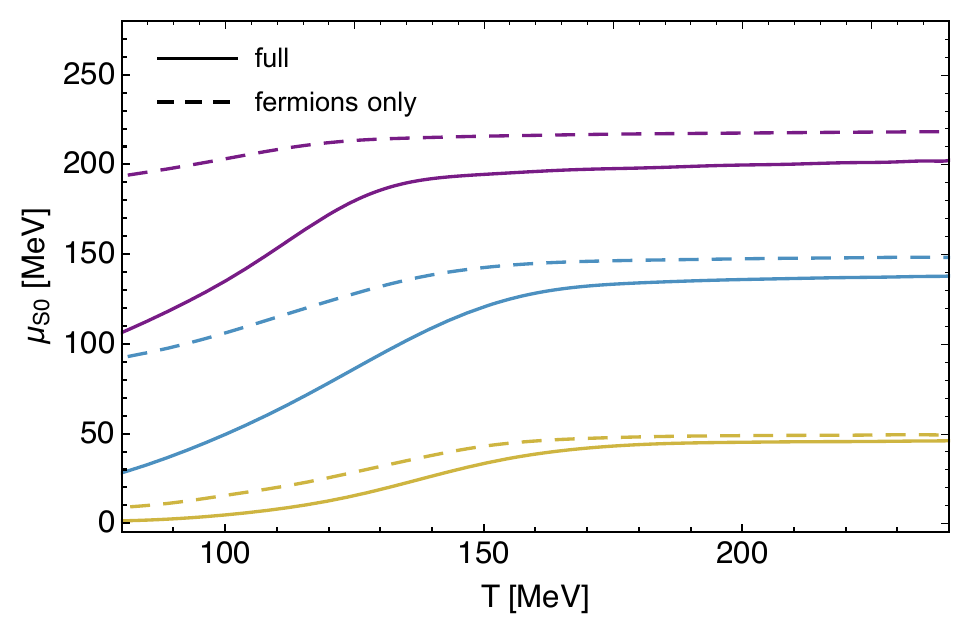}
\caption{Comparison between our full result for $\mu_{S0}$ (solid
  lines) and \Eq{eq:fuku} (dashed lines) for $\mu_B = 150,\, 450$ and
  $660$ MeV (from bottom to top). The color coding for $\mu_B$ is the
  same as in \Fig{fig:musovert}.}
\label{fig:fukurel}
\end{figure}

We show a comparison between our
full result for $\mu_{S0}$ and \Eq{eq:fuku} in \Fig{fig:fukurel}. We
have used the loops computed at $\mu_S = 0$ in this figure but have
checked that the results depend only very mildly on this choice. While
$L$ and $\bar L$ show a considerable dependence on $\mu_S$, their
ratio does not, excluding the former explanation for possible
deviations. We see that \Eq{eq:fuku} captures the qualitative trend of
$\mu_{S0}$ quite well, but is quantitatively very inaccurate. At
temperatures below the phase transition the difference can be
attributed to the missing effect of open strange mesons in
\Eq{eq:fuku}. This highlights the crucial importance of meson
fluctuations for strangeness neutrality. At larger temperatures the
asymptotic value $\mu_{S0} = \mu_B/3$ is rapidly reached with
\Eq{eq:fuku}. The reason is that $\bar L / L \approx 1$ in this case,
even though they are still smaller than one. As argued above, in our
full result the asymptotic value is not reached since the system is in
the semi-QGP phase. The heuristic relation does not capture this
feature at all. We want to emphasize that $\bar L / L \rightarrow 1$
at large $T$ crucially depends on the parametrization of the Polyakov
loop potential. In our case, \Eq{eq:plpot}, the Haar measure of the
gauge group is implemented directly into the potential. This restricts
the values of the loops to $L,\bar L \in [0,1]$. For different
parametrizations without the Haar measure the ordering $\bar L > L$ at
finite $\mu_B$ persists for arbitrarily large temperatures, with loops
larger than one. In this case, \label{eq:fukurel} would also yield
$\mu_{S0} < \mu_B/3$ at large $T$.

\subsection{Strangeness neutrality and QCD thermodynamics}\label{sec:td}

\begin{figure*}[t]
\centering
\includegraphics[width=.32\textwidth]{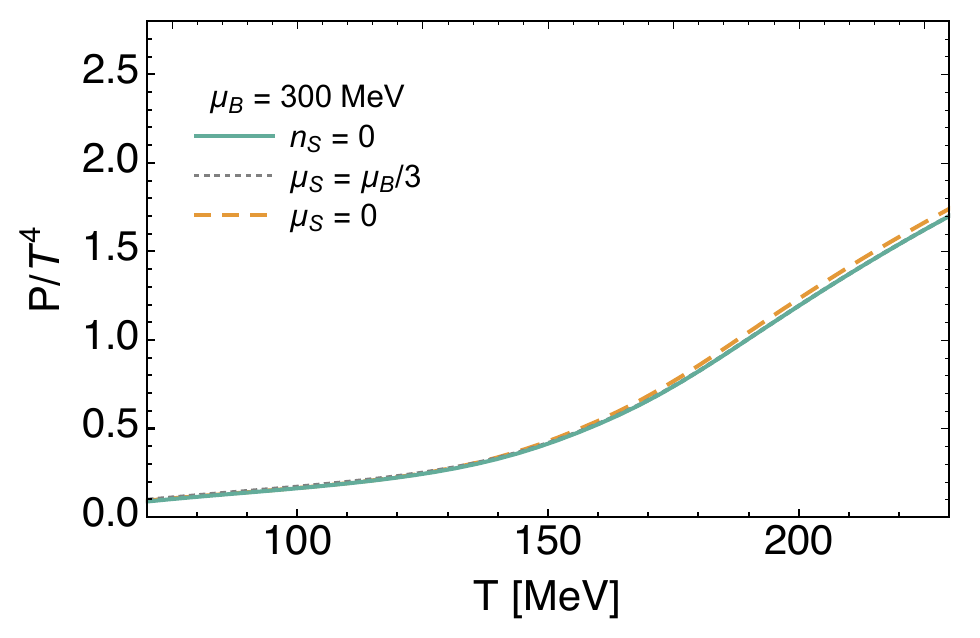}
\includegraphics[width=.32\textwidth]{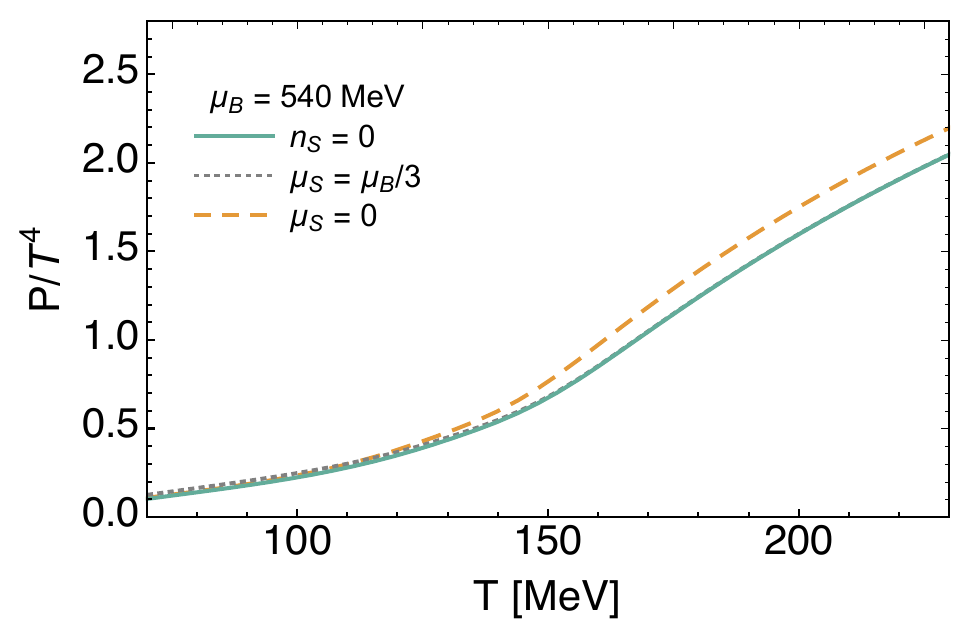}
\includegraphics[width=.32\textwidth]{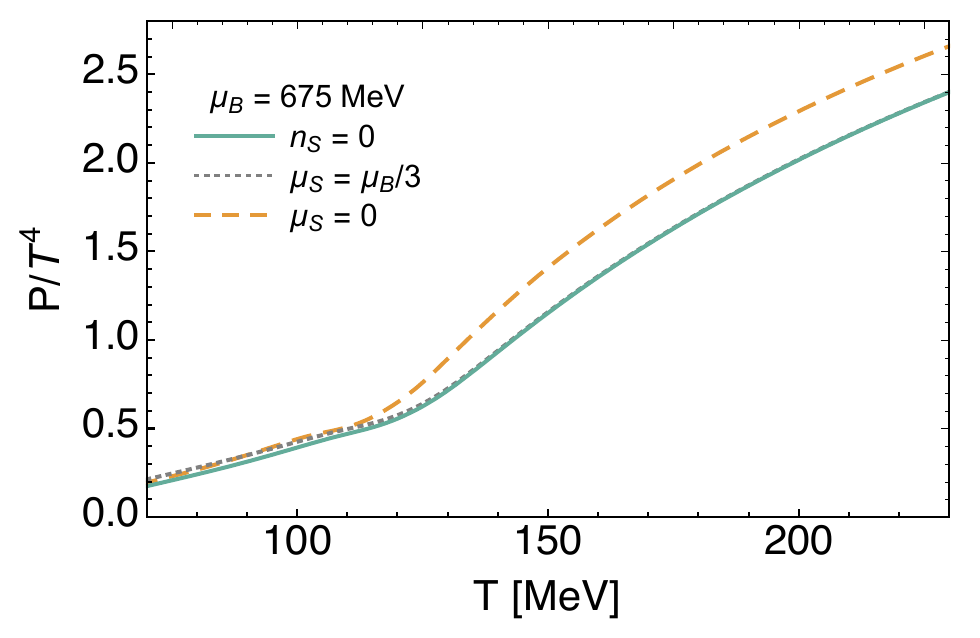}
\includegraphics[width=.32\textwidth]{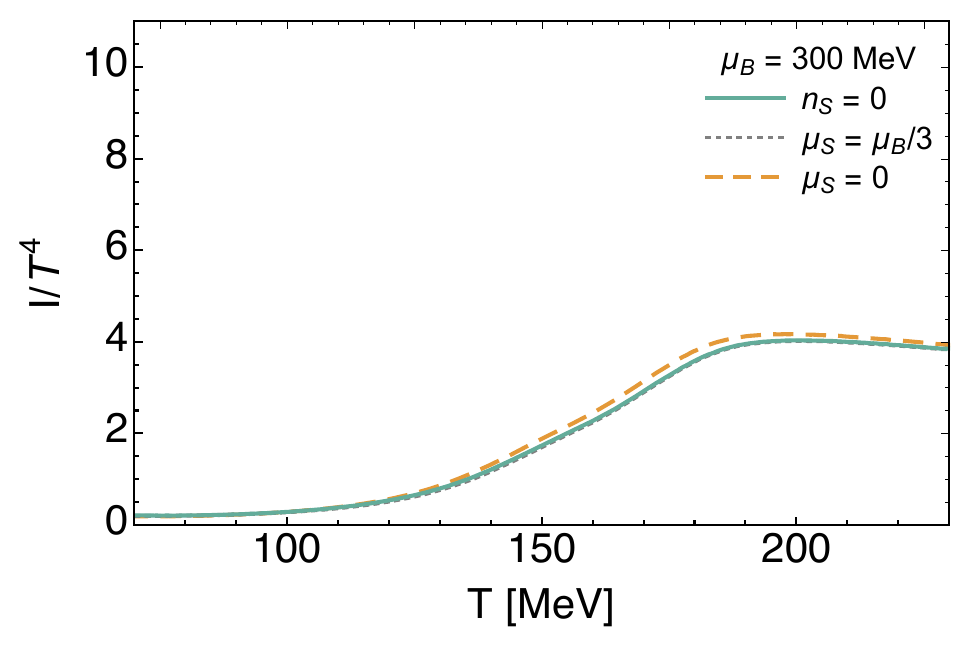}
\includegraphics[width=.32\textwidth]{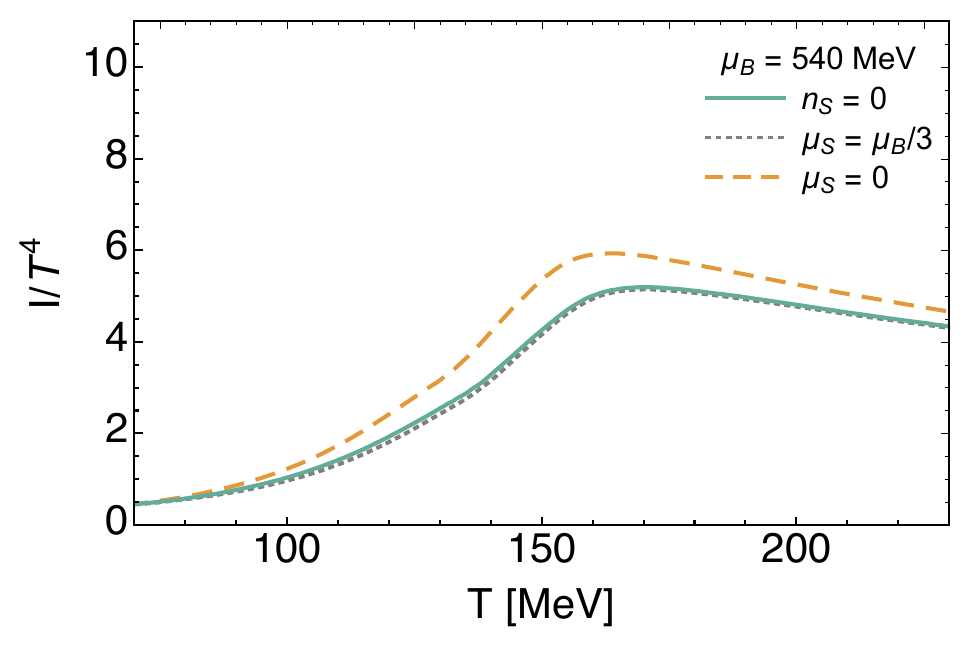}
\includegraphics[width=.32\textwidth]{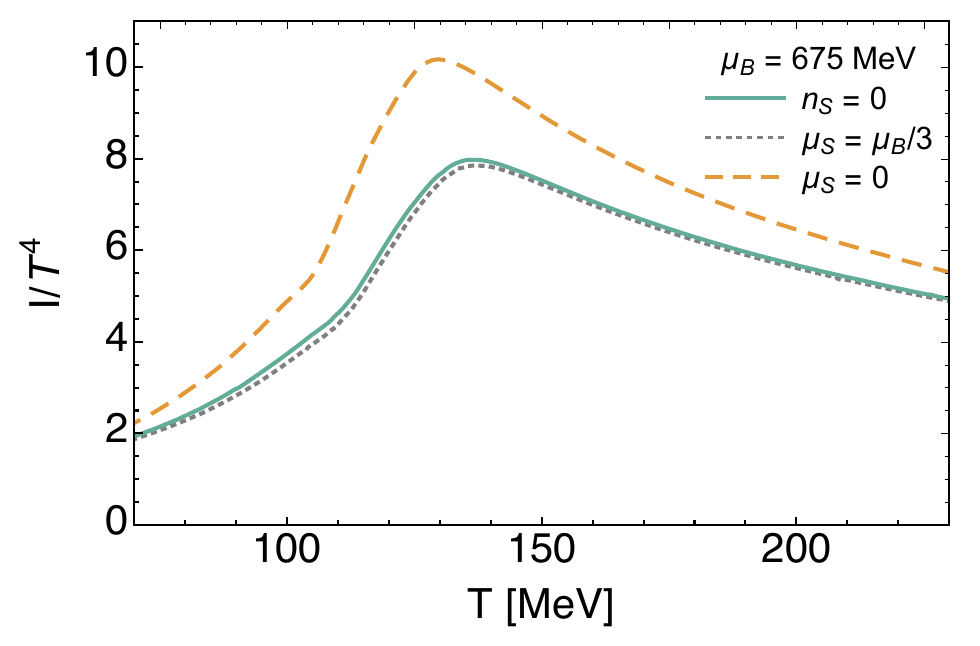}
\includegraphics[width=.32\textwidth]{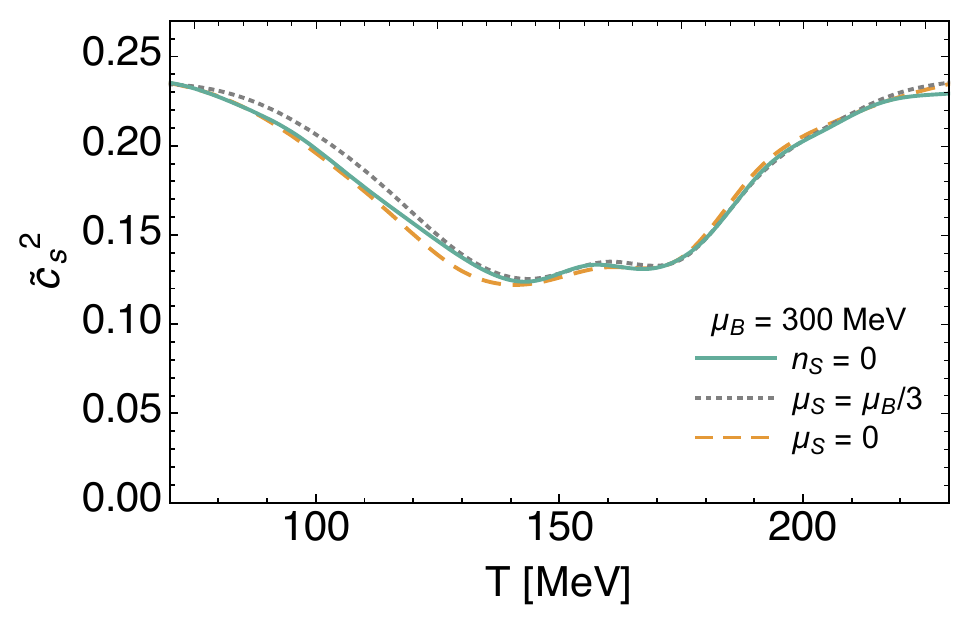}
\includegraphics[width=.32\textwidth]{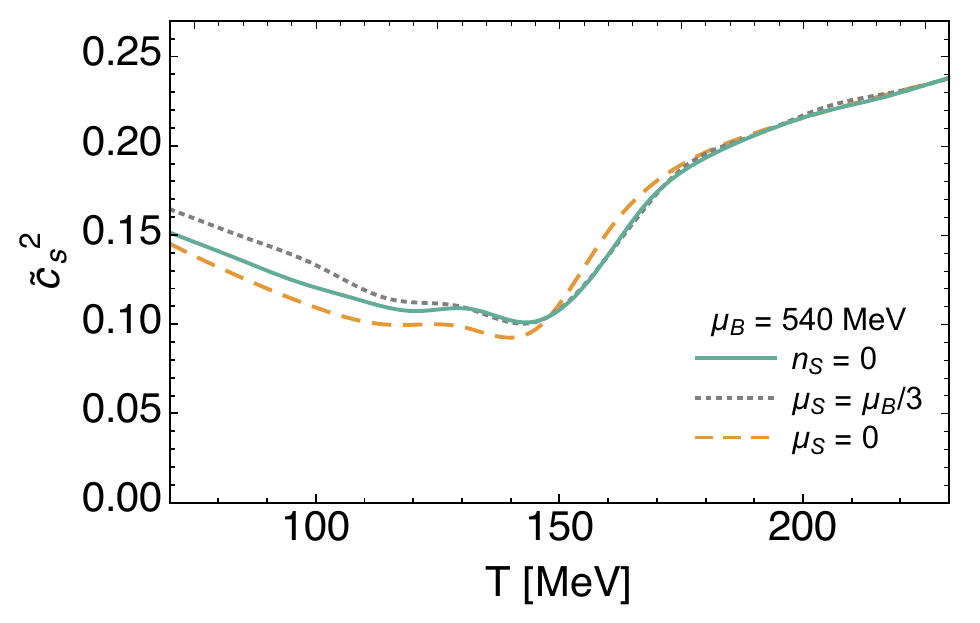}
\includegraphics[width=.32\textwidth]{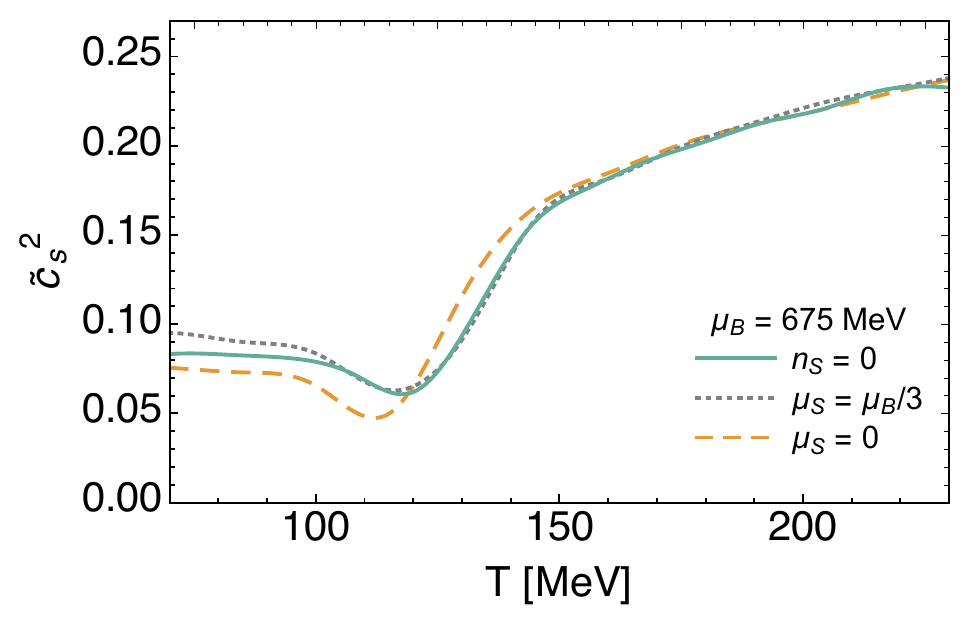}
\caption{Comparison between the pressure (first row), the trace anomaly (second row) and the speed of sound squared (third row) at strangeness neutrality (solid blue line), at $\mu_S = 0$ (dashed orange line) and at $\mu_S = \mu_B/3$ (dotted gray line) for various $\mu_B$; see \Eq{eq:thermodefs} for the definitions of these quantities.}
\label{fig:tdcomp}
\end{figure*}

We can now use the results of the previous section to investigate the influence of the strangeness neutrality on thermodynamic quantities. To this end, we compare our results at $\mu_S = 0$ (dashed, orange) and $\mu_S = \mu_B/3$ (dotted, gray) to the ones at strangeness neutrality, $n_S = 0$ (solid, green), at various $\mu_B$. This is shown in \Fig{fig:tdcomp}. The first row shows the pressure, the second the trace anomaly and the third the speed of sound squared. For small baryon chemical potential, $\mu_B \lesssim 300$ MeV, the equation of state is not very sensitive to the chemical potentials since baryon excitations are highly suppressed. At small temperatures pion fluctuations dominate the equation of state in this case and hence the thermodynamic quantities are essentially independent of $\mu_S$. At larger temperatures we find that the pressure and the trace anomaly are always smaller at strangeness neutrality than at $\mu_S = 0$. At larger $\mu_B$ this effect is more pronounced. The pressure and the trance anomaly start to grow at larger $T$ at strangeness neutrality as compared to $\mu_S = 0$, indicating that the QCD phase transition is shifted to larger temperatures. This is also apparent from the position of the minima of $c_s^2$, which approximately coincide with the pseudocritical deconfinement and chiral transition temperatures. Note that at $\mu_B = 675$ MeV we find $T_d \approx T_\chi$, so the two corresponding minima are degenerate. For $\mu_B = 675$ MeV the equation of state shows a sizable dependence on the strangeness. For the pressure we find a difference of about 20\% between $\mu_S = 0$ and $n_S = 0$ at large temperatures and for the the trace anomaly even more than 35\% in the transition region. The higher sensitivity of the trace anomaly is due to its direct dependence on the particle numbers. At strangeness neutrality, the baryon number is always smaller than at $\mu_S = 0$ for finite $\mu_B$ for all temperatures. This is as expected since finite $\mu_S$ leads to less strange particles in the system that can contribute to the baryon number.

In contrast to $p$ and $I$, the speed of sound squared shows the highest sensitivity in the small and intermediate temperature region. As discussed in \Sec{sec:tdmu0}, $p$ and $I$ are dominated by the increase in the number of degrees of freedom at the phase transition, while $c_s^2$ is not. In the hadronic regime we find a difference of about 30\% between $\mu_S = 0$ and $n_S = 0$ at $\mu_B = 675$ MeV. This is also apparent from the comparison to the results at $\mu_S = \mu_B/3$. As argued in the previous section, $\mu_S = \mu_B/3$ enforces strangeness neutrality in case of uncorrelated quarks, i.e.~deep in the deconfined phase. The results for $\mu_S = \mu_B/3$ and $n_S = 0$ should therefore become degenerate at large temperatures. This is also what we observe for the thermodynamic quantities. Since $\mu_{S0}$ is already close to its asymptotic value at $T_\chi$, cf.~\Fig{fig:musovert}, they are already very similar close to the chiral transition for $\mu_S = \mu_B/3$ and $n_S = 0$. The pressure and the trace anomaly show only very small differences between $\mu_S = \mu_B/3$ and $n_S = 0$ at small temperatures. $c_s^2$ shows a stronger sensitivity to the strangeness below the chiral phase transition. $\mu_S = \mu_B/3$ results in a larger and $\mu_S = 0$ in a smaller speed of sound in the hadronic phase as compared to the result at strangeness neutrality. This ordering is inverted for the pressure and the trace anomaly.

Overall, we found that the equation of state becomes increasingly sensitive to strangeness with increasing baryon chemical potential. At $\mu_B = 675$ MeV, where the transition is still a crossover in our model, the effects of strangeness neutrality as compared to vanishing strange chemical potential become as large as about 30\%.

\subsection{Strangeness neutrality and the phase structure}\label{sec:ps}

\begin{figure*}
\includegraphics[width=.885\columnwidth]{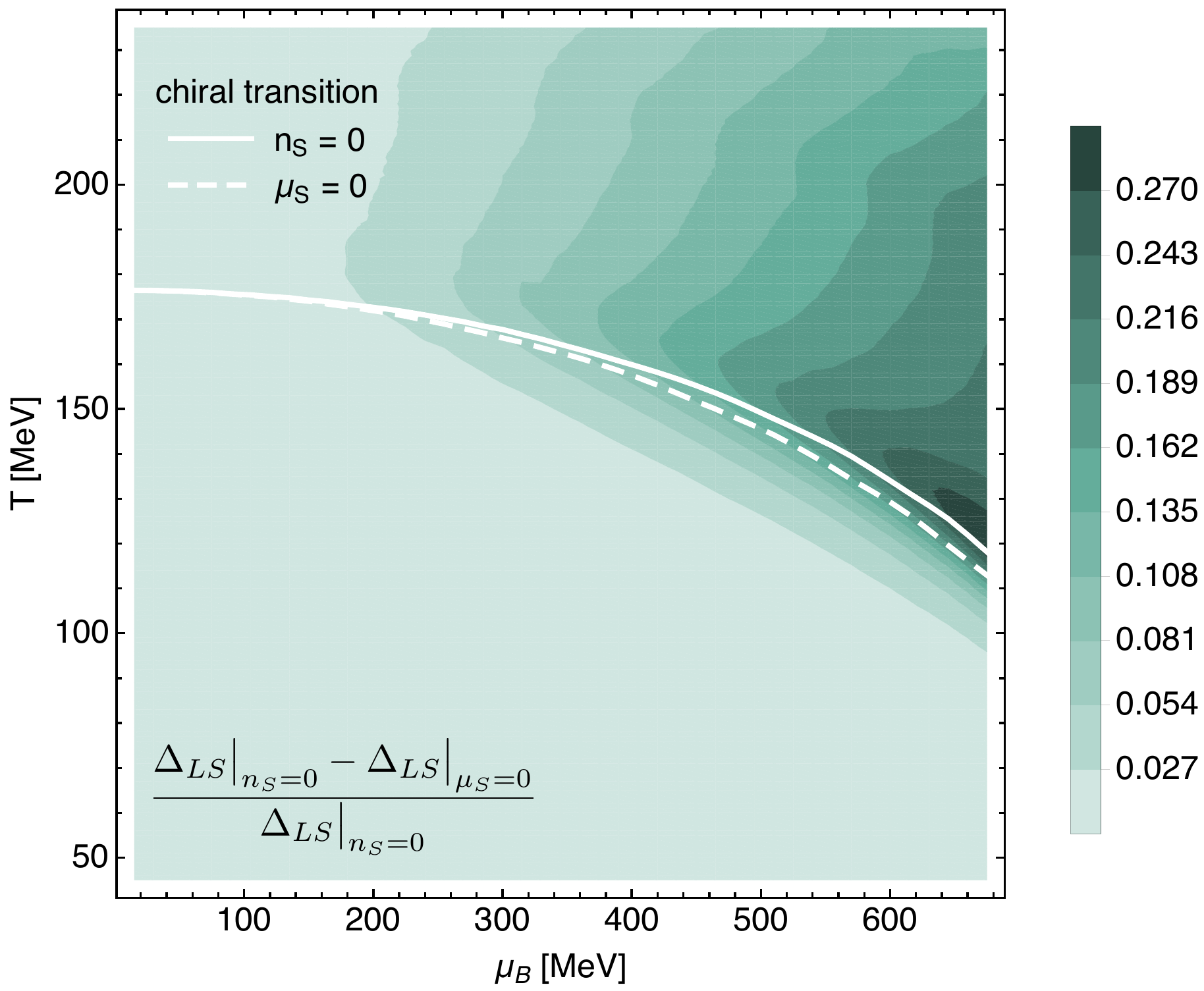}
\hspace{.9cm}
\includegraphics[width=.9 \columnwidth]{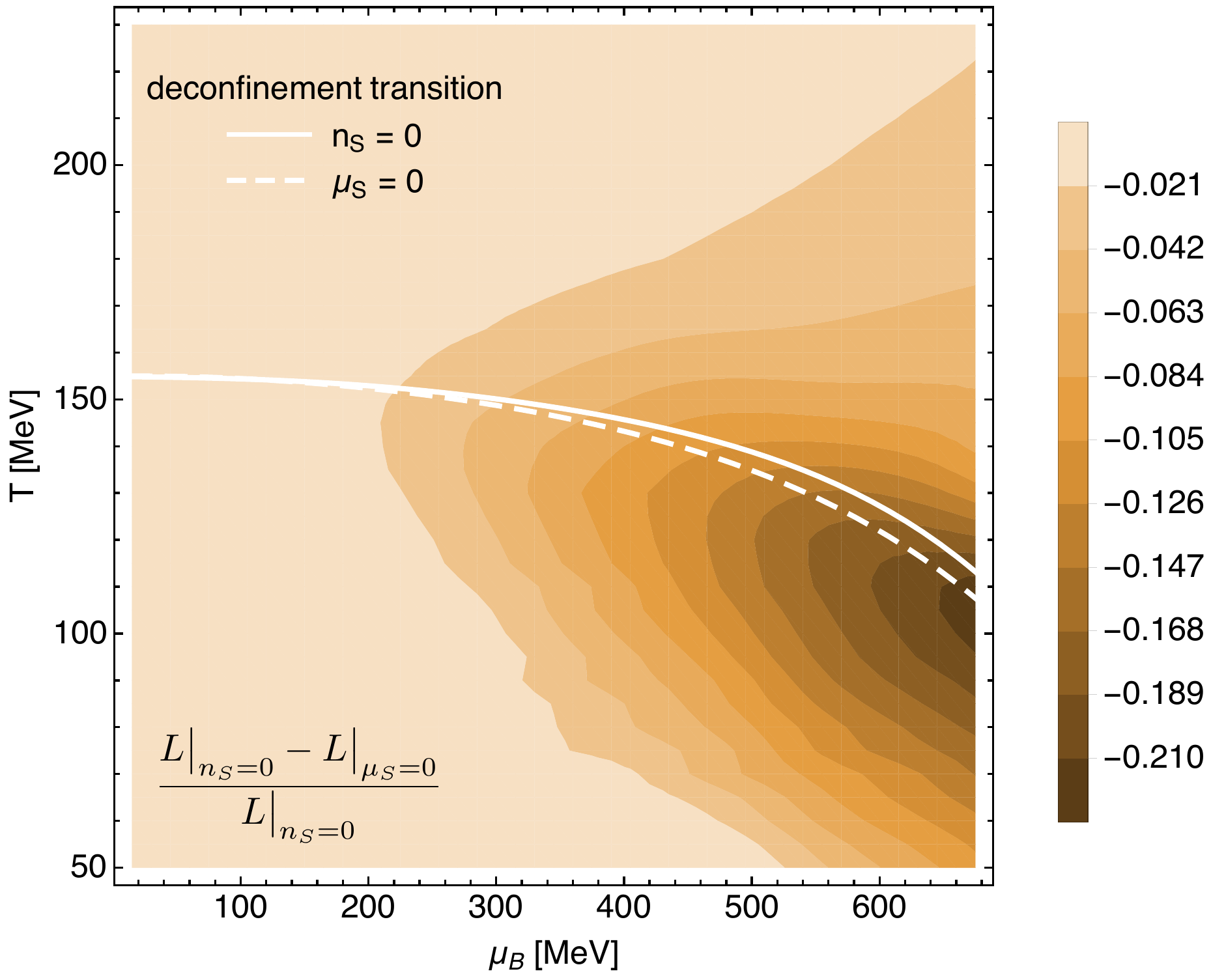}
\caption{Left: Relative error of the subtracted condensate for strangeness neutrality and $\mu_S = 0$. The solid  and dashed lines indicate the chiral phase boundary as defined by the inflection point of $\Delta_{LS}(T)$ for $n_S=0$ and $\mu_S=0$ respectively. Right: The same for the Polyakov loop. Here, the solid and dashed lines indicate the deconfinement phase boundary as defined by the inflection point of $L(T)$ for $n_S=0$ and $\mu_S=0$.}
\label{fig:pb}
\end{figure*}

As already indicated by the results in the previous section,
strangeness has a sizable effect on the phase structure at finite
baryon chemical potential. In the left plot of \Fig{fig:pb} we show
the phase diagram of the chiral transition as defined by the
inflection point of the subtracted chiral condensate, \Eq{eq:subcond},
at strangeness neutrality (solid line) and at vanishing strangeness
chemical potential (dashed line). We see, as already concluded in the
previous section, that strangeness neutrality leads to a larger
critical temperature as compared to $\mu_S=0$. The effect increases
with increasing $\mu_B$, but leads to only about 6\% difference in $T_\chi$
at the largest baryon chemical potential and is therefore very small. However, since the transition is a
crossover for the parameters considered here, it is more sensible to
compare the global structure of the order parameters. To this end, we
computed the relative difference between the subtracted condensate at
strangeness neutrality and at vanishing strange chemical potential,
\begin{align}\label{eq:scdiff}
\frac{\Delta_{LS}\big|_{n_S=0}-\Delta_{LS}\big|_{\mu_S=0}}{\Delta_{LS}\big|_{n_S=0}}\,.
\end{align}
The result is given by the density profile in the left plot of
\Fig{fig:pb}. The darker the color, the larger the difference. It
shows where the chiral phase structure is most sensitive to
strangeness. Similar to our findings for the pressure, the subtracted
chiral condensate is most sensitive at intermediate to large $\mu_B$
and above the critical temperature. In the hadronic phase, strangeness
neutrality does not have a big effect on the chiral order
parameter. Even though the effect of strangeness neutrality on the
inflection point of the order parameter is rather small, we find
deviations of up to about 27\% in the difference defined in
\Eq{eq:scdiff}. The relation
$\Delta_{LS}\big|_{n_S=0} \geq \Delta_{LS}\big|_{\mu_S=0}$ holds for
all $T$ and $\mu_B$ considered here.
Baryon effects (relative to meson effects), which tend to make the crossover steeper, are partly suppressed at strangeness neutrality since $\mu_S > 0$ effectively reduces strange baryon contributions. The chiral condensate therefore melts slower at strangeness neutrality.

A similar conclusion can be drawn for the deconfinement transition. In
the right plot of \Fig{fig:pb} we show the deconfinement transition as
defined by the inflection point of the Polyakov loop, \Eq{eq:poldef},
at strangeness neutrality (solid line) and at vanishing strange
chemical potential (dashed line). The antiloop $\bar L$ gives
essentially the same critical temperature. As for the chiral
transition, the pseudocritical temperature becomes slightly larger at $n_S = 0$
as compared to $\mu_S = 0$, where the difference increases with
increasing $\mu_B$. We also computed the relative difference
\begin{align}\label{eq:pldiff}
\frac{L\big|_{n_S=0}-L\big|_{\mu_S=0}}{L\big|_{n_S=0}}\,,
\end{align}
and the result is given by the density profile in the right plot of
\Fig{fig:pb}. Again, we find that the deviation grows with $\mu_B$ but
this time is largest in the hadronic regime right below the phase
boundary. Recalling that the deconfined phase corresponds to chiral
symmetry \emph{restoration} and center symmetry \emph{breaking}, we
conclude that both for the chiral and the deconfinement order
parameter, the transition region at large $\mu_B$ towards the
respective symmetry \emph{restored} phase is most sensitive to
strangeness. For the Polyakov loops we always find
$L\big|_{n_S=0} \leq L\big|_{\mu_S=0}$. The overall effect on the
deconfinement transition is a bit smaller than on the chiral
transition, but still about 20\%. These findings might suggest that
the results for the effect of strangeness neutrality on the
thermodynamic quantities in the previous section could be attributed
to the pressure and the trance anomaly being more sensitive to the
chiral transition, while the speed of sound is more sensitive to the
deconfinement transition.

Finally, we studied how strangeness neutrality affects the isentropes
in the phase diagram. They are defined by trajectories of constant
$s/n_B$. Without dissipation, i.e.\ the ideal case, the hydrodynamic
evolution of the quark-gluon plasma is along such isentropes. This is
due to the fact that without dissipation and only strong interactions,
both the entropy density and the baryon number are conserved in the
hydro evolution. Even though it is established by now that the QGP is
not an inviscid fluid, given the small shear viscosity over entropy
density of the QGP suggest by hydrodynamic simulations of heavy-ion
collisions, the isentropes still provide a good estimate for the
approximate path that the QGP in its late stages takes through the
phase diagram.

\begin{figure}[t]
\centering
\includegraphics[width=\columnwidth]{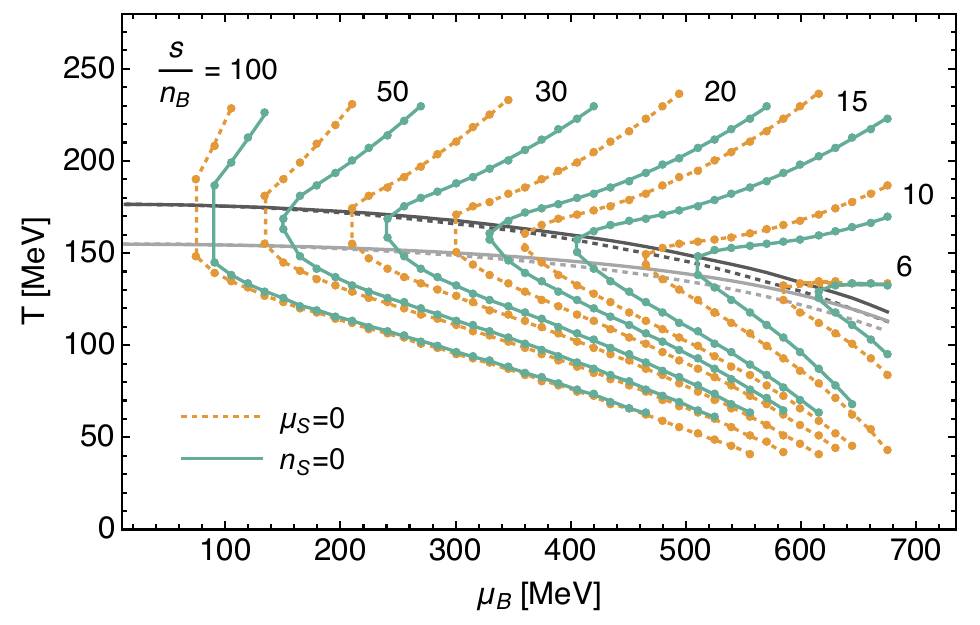}
\caption{Isentropes in the phase diagram. The dark and light gray lines are the chiral and deconfinement phase boundaries respectively.}
\label{fig:isen}
\end{figure}

Our results are shown in \Fig{fig:isen}. The orange dashed line
corresponds to $\mu_S = 0$ and the solid blue line shows the
isentropes at strangeness neutralities for various fixed ratios
$s/n_B$. The isentropes show a very characteristic behavior: they have
positive slope in the phase diagram above the phase transition and a
negative slope below. In the transition region, the slope changes
sign, with a slower `turning' of the isentropes at smaller $\mu_B$,
where the crossover region is wider. We find this kink even at large $s/n_B$. Interestingly, in studies of the isentropes within two-flavor QM and PQM models such a kink only occurs for small $s/n_B$ \cite{Nakano:2009ps, Skokov:2010uh}. Hence, the sensitivity of the isentropes to the phase transition at large $s/n_B$ can be attributed to srangeness.

The behavior of the isentropes in
the hadronic phase is dictated by the Silver-Blaze property of QCD. At
$T = 0$ and $\mu_B \lesssim 3 m_l$ the baryon number has to
vanish. Hence, the isentropic curves bend toward larger $\mu_B$ with
decreasing $T$. The difference between $n_S = 0$ and $\mu_S=0$ is
small at small temperatures because the lightest baryonic resonance
does not carry strangeness. Since the system is in the semi-QGP phase
above the phase transition, the entropy density has not reached its
asymptotic value yet and is hence still growing with $T$. The baryon
number, on the other hand, has a maximum at the chiral phase
transition and slowly decreases with increasing temperature above
$T_\chi$. Hence, the isentropes bend towards larger $\mu_B$ with
increasing $T$ above the phase transition. The regions where the
isentropes turn therefore clearly indicate the transition
region. Since the baryon number at strangeness neutrality is
systematically smaller than for $\mu_S = 0$ at a given $\mu_B$, the
bending of the isentropes above the phase transition is stronger at
strangeness neutrality. We also find that the isentropes at
strangeness neutrality are systematically shifted to the
right. Qualitatively, this can be understood from the fact that the
baryon number decreases with increasing $\mu_S$. This effect dominates
over the corresponding effect on the entropy density (which behaves
very similar to the pressure in \Fig{fig:tdcomp}). Thus, larger
$\mu_B$ is necessary to ensure a fixed $s/n_B$ at strangeness
neutrality.

\section{Summary}\label{sec:sum}

Strangeness neutrality is a crucial property of the matter created in
heavy-ion collisions. We studied its impact on
QCD thermodynamics and the phase structure. To this end, we set up a
2+1 Polyakov loop enhanced quark-meson model that captures the
dynamics of mesons, quarks and, to some extent, baryons in a gluon
background field at finite baryon and strangeness chemical
potential. We demonstrated by comparing to available lattice data that
this works very well for the QCD equation of state not only at
vanishing chemical potential, but also at finite $\mu_B/T$.

Demanding that the strangeness number is always zero implicitly
defines a corresponding strange chemical potential as a function of
temperature and baryon chemical potential. We computed resulting
function $\mu_{S0}(T,\mu_B)$. Its non-trivial functional form has a
transparent interpretation in terms of competing strange meson and
baryon dynamics at finite baryon chemical potential and is therefore
intimately tied to confinement. We compared these results to the
purely fermionic case, i.e.\ where only quark and baryon dynamics are
taken into account, and found huge discrepancies. This highlights the
crucial importance of open strange meson dynamics for the accurate
description of strangeness physics and the freeze-out conditions of
heavy-ion collisions.

We used our results for $\mu_{S0}(T,\mu_B)$ to compute QCD
thermodynamics and the phase structure at strangeness neutrality. The
effect of the strangeness content of the QCD medium on its
thermodynamics is certainly interesting on its own right, but also
very important as an input for, e.g., the hydrodynamic description of
heavy-ion collisions. The comparison of our results at vanishing
density to lattice QCD results show very good agreement, even for the
highly non-trivial speed of sound. To assess the effect of strangeness
neutrality we confronted results on the equation of state at fixed
strange chemical potential, where we have chosen $\mu_S = 0$ and
$\mu_B/3$, to the equation of state at strangeness neutrality. For
reasons related to the range of validity of our model (see
\App{app:tdlamu}) we restricted our analysis to
$\mu_B \in \{0,\dots,675\}$ MeV but note that this covers the region
probed by current beam energy scan experiments \cite{Alba:2014eba}
(assuming that the translation of the beam energy to the baryon
chemical potential based on the hadron resonance gas is correct). Our
results show that the relevance of strangeness neutrality grows with
increasing baryon chemical potential and the difference between
strangeness neutrality and $\mu_S = 0$ can be as large as about 30\%
at $\mu_B = 675$ MeV, in particular for the trace anomaly and the
speed of sound squared.

We find a similar sensitivity of the chiral and deconfinement phase
transitions on strangeness. Overall, the pseudocritical temperatures
of both transition are larger at strangeness neutrality than at vanish
strange chemical potential. Hence, strangeness neutrality `delays' the transition to the QGP. Again, while the effect is small at small
$\mu_B$ and becomes considerable at larger $\mu_B$. This can be
attributed to a suppression of symmetry-breaking fermionic
fluctuations in the strange sector due to finite $\mu_S$. Due to their
distinct sensitivity to the phases of QCD and the related
thermodynamics, the isentropes, which provide a good estimate for the
path of the hydrodynamic evolution of the QGP though the phase
diagram, also turned out to be affected by strangeness neutrality
significantly.

In summary, we have demonstrated that the QCD equation of state and
its phase structure are highly sensitive to the strangeness content of
the medium. For the accurate description of heavy-ion collisions at
varying beam energies it is indispensable to take this into account. The
underlying physics is very intriguing since the strangeness neutrality
condition $n_S = 0$ is sensitive to various characteristic properties
of QCD, namely the interplay of meson and baryon dynamics at finite
chemical potential as well as the chiral and deconfinement phase
structure. The present results facilitate the computation of
fluctuation observables in heavy-ion collisions, such as higher
cumulants of baryon number and strangeness distributions including
off-diagonal cumulants, under more realistic conditions.

Towards a more realistic equation of state, the next crucial step is
to also account for the freeze-out condition related to the initial
charge of the colliding nuclei by taking finite isospin chemical
potential into consideration. Also in this case, beyond mean-field
effects and in particular pion fluctuations will certainly be very
important. Concerning the model, the most relevant improvements are
the inclusion of effects beyond LPA which have a high impact on quark
and meson dynamics, and the incorporation of dynamics in the gauge
sector which allow for a self-consistent computation of the Polyakov
loop potential. The latter point might remedy the
thermodynamic inconsistency of the PQM model at large $\mu_B$
discussed in the appendix and thus allow for an extension of the
present work towards the critical endpoint of QCD. Then,
(off-diagonal) cumulants of baryon number and strangeness
distributions will also become accessible.


\vspace{0.1cm} {\it Acknowledgments -} We thank Mario Mitter, Robert
D. Pisarski, Vladimir Skokov and Patrick Steinbrecher for 
discussions as well as Jens Braun and Bernd-Jochen Schaefer for valuable comments on our manuscript. We also thank the members of the fQCD collaboration \cite{fQCD} for discussions and work on related projects. We thank Dirk H.\ Rischke for helpful explanations regarding the speed of sound. F.R. is supported by the Deutsche Forschungsgemeinschaft
(DFG) through grant \mbox{RE 4174/1-1}.  W.F. is supported by the National Natural Science Foundation of China under Contracts Nos. 11775041.  This work is supported by the ExtreMe Matter Institute
(EMMI) and the grant BMBF 05P12VHCTG.  It is part of and supported by
the DFG Collaborative Research Centre "SFB 1225 (ISOQUANT)".

\begin{appendix}

\section{Details on the Polyakov loop potential}\label{app:poloop}

Here we provide the details on the Polyakov loop potential $U_\text{glue}$. For $x=(a,c,d)$ the temperature dependent coefficients in \Eq{eq:plpot} are of the form
\begin{align}\label{eq:plpotcoeffs}
\begin{split}
x(T) &= \frac{x_1+x_2/t+x_3/t^2}{1+x_4/t+x_5/t^2}\,,\\
b(T) &= b_1\, t^{-b_4} \big(1-e^{b_2/t^{b_3}} \big)\,,
\end{split}
\end{align}
where the parameters have been determined in \cite{Lo:2013hla} and are shown in \Tab{tab:plparas}. $t = t_\text{red}+1$ with $t_\text{red}= \alpha_t (T-T_0)/T_0$. $T_0$ is the deconfinement temperature of the pure gauge theory, while $\alpha_t$ is a parameter that controls the speed of the transition. Due to unquenching effects, 
both parameters deviate from the values of the pure gauge theory, $T_{0,\text{YM}} = 276$ MeV and $\alpha_{t,\text{YM}} = 1$. Since the QCD transition has a smaller critical temperature and a smoother transition, one generally expects $T_0<T_{0,\text{YM}}$ and $\alpha_t < \alpha_{t,\text{YM}}$. In \cite{Haas:2013qwp} $\alpha_t = 0.57$ has been determined. However, since this depends on the number of flavors, the truncation and the parametrization of the Polyakov loop potential, we will consider both $\alpha_t$ and $T_0$ as free parameters here. They can be determined, e.g., by fitting the pressure to the lattice result at vanishing density. The other fit parameters of the potential are given by their YM values and are given in \Tab{tab:plparas}:

\begin{table}[t]
\begin{center}
\begin{tabular}{ c | c | c | c | c }
\hline \hline
$a_1$ & $a_2$ & $a_3$ & $a_4$ & $a_5$ \\
\hline
 -44.14 & 151.4 & -90.0677 & 2.77173 & 3.56403 \\
\hline \hline
$b_1$ & $b_2$ & $b_3$ & $b_4$ & $ \phantom{b_4}$ \\
\hline
-0.32665 & -82.9823 & 3.0 & 5.85559 & \phantom{} \\
\hline \hline
$c_1$ & $c_2$ & $c_3$ & $c_4$ & $c_5$ \\
\hline
-50.7961 & 114.038 & -89.4596 & 3.08718 & 6.72812 \\
\hline \hline
$d_1$ & $d_2$ & $d_3$ & $d_4$ & $d_5$ \\
\hline
 27.0885 & -56.0859 & 71.2225 & 2.9715 & 6.61433 \\
\hline \hline
\end{tabular}
\end{center}
\caption{Fit parameters of the Poyakov loop potential defined in Eqs.~(\ref{eq:plpot}) and (\ref{eq:plpotcoeffs}). These are taken from \cite{Lo:2013hla}.}
\label{tab:plparas}
\end{table}

The inclusion of finite chemical potentials to the gauge sector can be
achieved along the lines of \cite{Schaefer:2007pw, Herbst:2013ail}. It
is constructed phenomenologically from the identification of
$\Lambda_\text{QCD}$ in the one-loop beta function of QCD at large
density (HTL/HDL) with the flavor dependent modification of the
critical temperature. This suggests the following modification of
$T_0$, \cite{Schaefer:2007pw},
\begin{align}
T_0(N_f,\mu) = T_\tau e^{-1/(\alpha_0 b_\mu)}\,,
\end{align} 
where $T_\tau = 1.77$ GeV sets the renormalization scale with the
corresponding coupling $\alpha_0 = 0.304$ for $N_f = 0$. $b_\mu$
encodes the flavor and chemical potential dependence of beta function:
\begin{align}\label{eq:bmu}
b_\mu = b_0 + \frac{16}{\pi} \bigg[ 2\frac{\mu^2}{(\hat\gamma T_\tau)^2} \Delta n_l + \frac{(\mu-\mu_S)^2}{(\hat\gamma T_\tau)^2} \Delta n_s \bigg]\,.
\end{align}
$b_0$ can be chosen either to be the well-known one-loop QCD beta function coefficient, $b_0 = (11 N_c-2 N_f)/(6\pi)$, or, in the spirit of $T_0$ as an approximation dependent free parameter, to also be a free parameter. The second term in \Eq{eq:bmu} is constructed such that the chiral and deconfinement transition agree at finite $\mu$ at mean-field in the two flavor PQM \cite{Schaefer:2007pw}. $\hat\gamma$ can be used as an additional parameter to control the curvature of the deconfinement phase transition. We use $\hat\gamma=1$ for the time being. The distributions $\Delta n_{l/s}$ are introduced in order to maintain the Silver Blaze property at vanishing temperature. For $\Delta n_{l/s} = 1$ the above parametrization would yield a $\mu$-dependent equation of state at vanishing temperature. Under the requirement that $\Delta n_{l/s} \rightarrow \Theta(\mu-M_{l/s})$ at vanishing temperature, we define
\begin{align}
\begin{split}
\Delta n_l &= \frac{1}{e^{3(M_l-\mu)/T}+1} + \frac{1}{e^{3(M_l+\mu)/T}+1}\\
&\quad- \frac{2}{e^{3 M_l/T}+1}\,,\\
\Delta n_s &= \frac{1}{e^{3(M_s-\mu+\mu_S)/T}+1} + \frac{1}{e^{3(M_s+\mu-\mu_S)/T}+1}\\
&\quad- \frac{2}{e^{3 M_s/T}+1}\,,
\end{split}
\end{align}
where $M_{l/s}$ are renormalized vacuum masses of the light and strange quarks.

\section{Thermodynamics at large $\mu$}\label{app:tdlamu}

\begin{figure}[t]
\centering
\includegraphics[width=.47\textwidth]{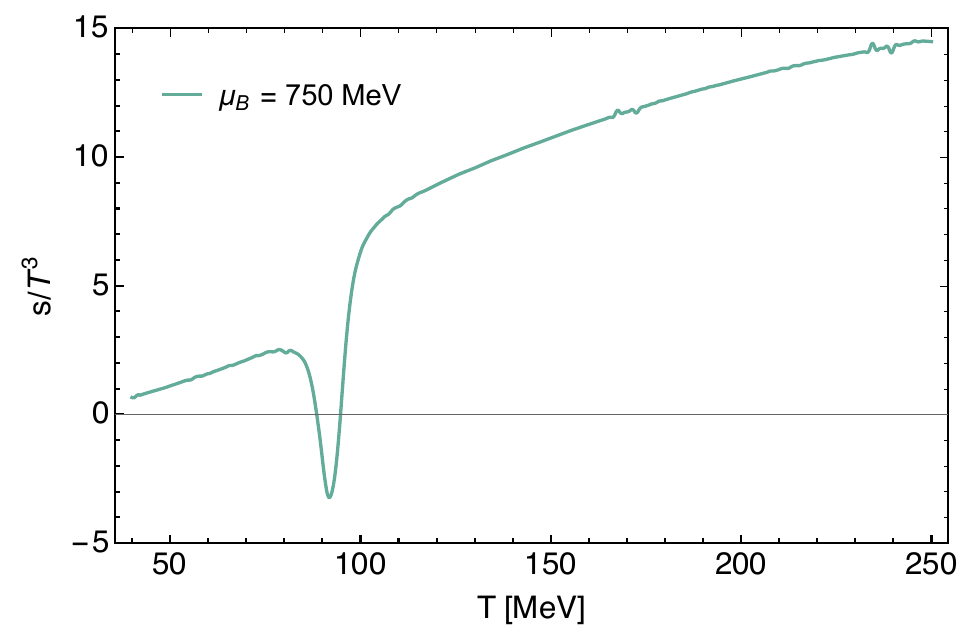}
\caption{Entropy density at $\mu_B = 750$ MeV and $\mu_S = 0$.}
\label{fig:smu250}
\end{figure}

\begin{figure}[t]
\centering
\includegraphics[width=.47\textwidth]{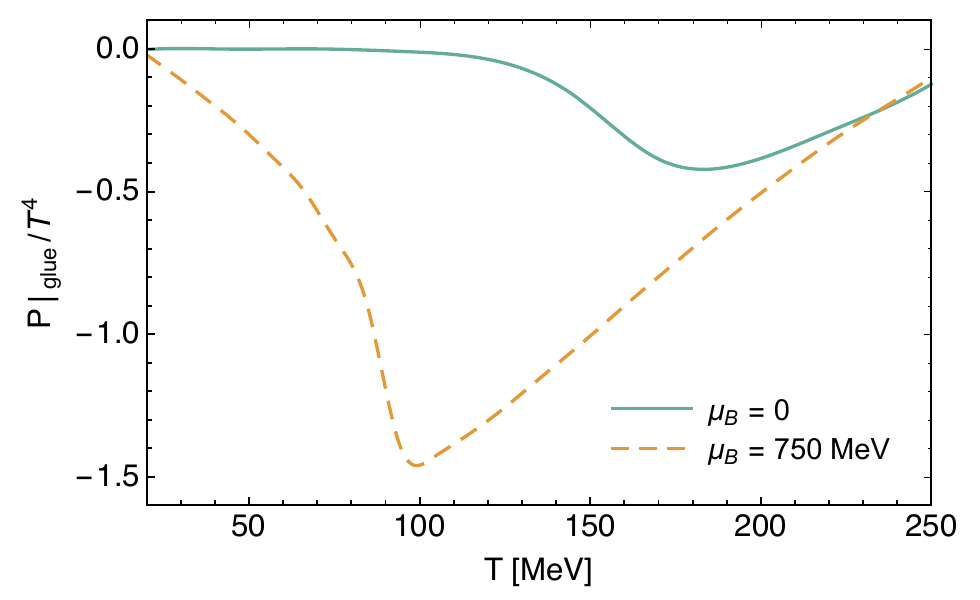}
\caption{Contribution of the Polyakov loop potential to the pressure.}
\label{fig:pglue}
\end{figure}

Throughout this work, we have used $\mu_B \leq 675$ MeV. This is below the critical endpoint of the model, which would certainly be interesting to study also in the context of this work. We find that starting at $\mu_B \gtrsim 700$ MeV the pressure develops an increasingly strong non-monotonoticity with increasing $\mu_B$ in the vicinity of the phase transition. This eventually leads to a negative entropy density in this region, as shown in \Fig{fig:smu250} at $\mu_B = 750$ MeV and $\mu_S = 0$. We explicitly checked that this is independent of the parametrization of the loop potential. The origin of this behavior can be traced back to the contribution of the gauge sector to the pressure,
\begin{align}
p\big|_\text{glue} = - U_\text{glue}(L,\bar L)\,,
\end{align}
where the Polyakov loops are part of the solutions of the equations of motion. We show this contribution at $\mu_B = 0$ and $\mu_B =750$ MeV for $\mu_S = 0$ in \Fig{fig:pglue}. This contribution is negative and has a minimum at around the chiral transition temperature. We see that the larger $\mu_B$ the larger this negative contribution tho the pressure becomes. For the baryon chemical potentials used in the main part of this work, where the pressure is always monotonously increasing, this negative contributions can be interpreted as the suppression of hadronic contributions to the pressure in the transition region due to deconfinement. This effect then is clearly overestimated at large $\mu_B$, leading to unphysical thermodynamics.

This problem originates in a combination of potential effects:

Firstly and most prominently, the construction of the Polyakov loop potential we use in this work, \Eq{eq:plpot}, is based on the pressure, the expectation values of the Polyakov loops and their two-point correlators \cite{Lo:2013hla}. This corresponds to Taylor expansion to second order of the potential about the minimum. The pressure is the value of the potential, the Polyakov loop expectation value determines the location of the minimum and the two-point correlator determines the curvature in the minimum. Further information on the global form of the potential comes from the temperature dependence of the parameters and the Haar measure of the loop. Evidently, this does not fully constrain the potential away from the Yang-Mills minimum. Moreover, the potential is best constrained for $L=\bar L$. The further away from the expansion point the potential has to evaluated, in particular for $L\neq \bar L$, the less constrained it is. This could be cured by either taking into account higher correlation functions of the loops in an extension of \cite{Lo:2013hla}, or by using a self-consistent $\bar A_0$-potential from the FRG \cite{Braun:2007bx, Braun:2009gm, Fister:2013bh}.

Secondly, the effect of matter fluctuations is only taken into account effectively by a simple quark flavor and chemical potential dependent rescaling of the potential as discussed in \App{app:poloop}. While this works well at small chemical potential, it might be too simple at large chemical potential. This problem could be cured by a self-consistent FRG computation as mentioned above.

Thirdly, for large chemical potentials and temperatures the initial conditions depend on these external parameters. Within the present approximation this is discussed in \Sec{sec:ini}. More generally, information from QCD at large energy scales are required, see e.g.\cite{Springer:2016cji}.

Lastly, for large chemical potentials it might be possible that the free energy is minimized by an inhomogeneous solution. Consequently, our solution on a homogenous background could potentially lead to a negative contribution to the pressure, see, e.g., \cite{Nickel:2009wj, Carignano:2014jla, Braun:2015fva, Tripolt:2017zgc} for studies within (P)NJL and QM models. Given the explicit analysis done below and due to the occurrence of this problem already at moderate chemical potential, this is unlikely to be the origin of the problem in the present case. Furthermore, by using a Fierz-complete basis for the four-quark interaction channels within a NJL model, it has been shown in \cite{Braun:2017srn, Braun:2018bik} that other channels, for instance isoscalar-vector and diquark channels, become relevant for the phase structure at finite baryon chemical potential. Since we only account for the scalar-pseudoscalar channel in this work (cf.~\Sec{sec:let}), we might miss some relevant effects at larger chemical potential.

\begin{figure}[t]
\centering
\includegraphics[width=.47 \textwidth]{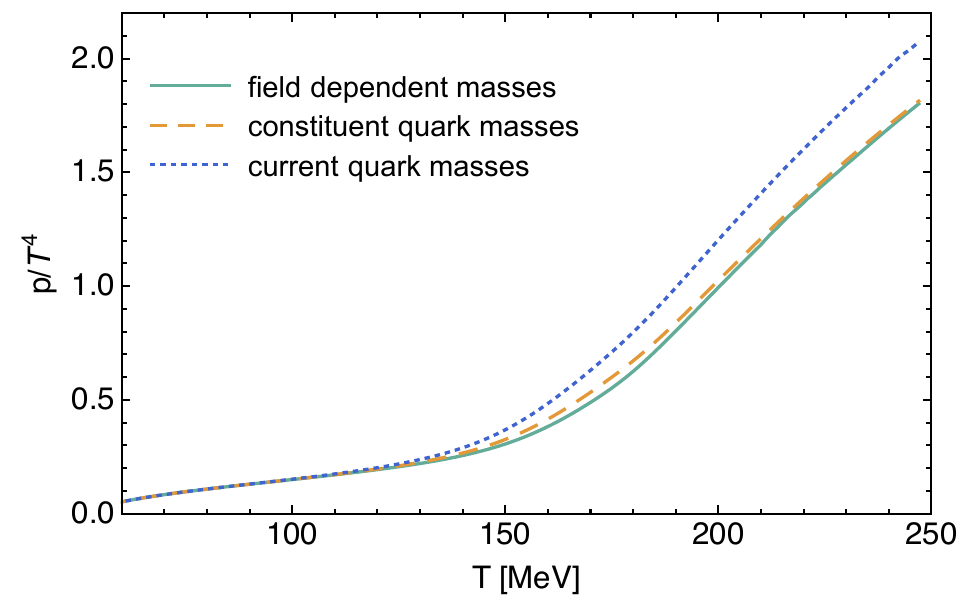}
\caption{The pressure at $\mu = 0$. The solid green line is computed with the full field dependence of $\Delta\Gamma_\Lambda$. The dotted blue line shows the results with $m_l = 3.6$ MeV and  $m_s =  95$ MeV. The dashed orange line correspond to $m_l = 300$ MeV and  $m_s =  430$ MeV.}
\label{fig:inifd}
\end{figure}

The problems discussed above manifest themselves in the gluon contribution to the pressure in the present work. In \Fig{fig:pglue} we show the contribution of $U_\text{glue}$ to the pressure at $\mu_B  = 0$ and $\mu_B  = 750$ MeV at vanishing strangeness chemical potential. Since the deconfinement transition in $SU(3)$ Yang-Mills theory is of first order, $U_\text{glue}$ is normalized such that its minimum is at zero for $T<T_0$. The Polyakov loops are also exactly zero in this case, $L^\text{YM} = \bar L^\text{YM} = 0$. In the present work, and Polyakov-loop enhanced models of QCD in general, the deconfinement transition is a crossover and the Polyakov loops $L,\,\bar L$ are always non-zero. This means that $U_\text{glue}$ is probed away from its normalized minimum, so while $U_\text{glue}(L^\text{YM},\bar L^\text{YM}) = 0$ for $T < T_0$ one has $U_\text{glue}(L,\bar L) > 0$. For increasing $\mu_B$ the Polyakov loops in QCD become larger and also unequal. So, as discussed above, we probe the potential in a region that is not well described by the present parametrization. This explains our observation in \Fig{fig:smu250} and why we refrain from doing computations at too large $\mu_B$.

\section{Field dependence of the initial conditions}\label{app:fdi}

Here, we check the effect of the meson field dependence of $\Delta\Gamma_\Lambda$ as discussed in \Sec{sec:ini}. One may argue that it is sufficient to use the current quark, or even vanishing, masses in \Eq{eq:medinitexpl} instead of resorting to background field dependent quark masses. However, it turns out that this is quantitatively very inaccurate in the present case. The medium-dependent corrections for the effective potential become relevant well before the quark masses reach their current values. In particular in the LPA, the quarks approach their current mass very slowly above $T_c$, if at all, so that they are reached well above the temperatures relevant here. As a result, setting $m_l \approx 3.6$ MeV and  $m_s \approx 95$ MeV leads to a significant overestimation of the in-medium corrections to the initial action. The same is true for the case where the initial quark masses that follow from the initial parameters in \Tab{tab:ini}, which are about a factor of 2-3 larger than the PDG current masses, are used.

However, it turns out that using the vacuum constituent quark masses, $m_l \approx 300$ MeV and $m_s \approx 430$ MeV works quite well. This is shown in \Fig{fig:inifd}. As argued in \Sec{sec:ini}, the most accurate determination of the equation of state is by using the background field dependent quark masses in the in-medium corrections of the initial conditions. This is the solid green line in the figure. Using the current quark masses leads to a considerable overestimation of the pressure, as shown by the dotted blue line. The dashed orange line shows the result with the constituent quark masses and we see that it gives a very accurate result. The error of this procedure is largest in the transition region, where it is about 8\%. We have checked explicitly that these findings are also true at finite chemical potentials. The advantage of the field independence is obviously that $\Delta\Gamma_\Lambda$ only enters in the initial pressure. Only $\omega_{00,\Lambda}$ in \Eq{eq:medinipot} receives a correction from $\Delta\Gamma_\Lambda$. The numerical integration of higher derivatives of \Eq{eq:medinitexpl} for the correction to the higher Taylor coefficients becomes unnecessary and irrelevant operators can be se to zero at the initial scale. At order $\phi^{10}$ this results in a speed-up by a factor of two to three with the numerical integration we implemented. Hence, given this large numerical speedup we accept the relatively small systematic error in our results on thermodynamics.

We would like to emphasize that these results apply to the fixed background Taylor expansion we used to solve the flow equation of the effective potential \cite{Pawlowski:2014zaa, Rennecke:2016tkm} and might not be directly transferrable to other methods. This is due to the fact that we expand the effective potential about its temperature and chemical potential dependent IR minimum. Using the current quark masses for the in-medium corrections of the initial effective potential is therefore consistent with our expansion scheme. For a more general discussion on this matter we refer to \cite{Braun:2018svj}.

\end{appendix}


\bibliography{qcd-phase2}

\end{document}